# Impact of O concentration on the thermal stability and decomposition mechanism of (Cr,Al)N compared to (Ti,Al)N thin films

**Authors:** Pauline Kümmerl[1*], Ganesh Kumar Nayak[1], Felix Leinenbach[1], Zsolt Czigány[2], Daniel Primetzhofer[3], Szilárd Kolozsvári[4], Peter Polcik[4], Marcus Hans[1], and Jochen M. Schneider[1]

[1] *Materials Chemistry, RWTH Aachen University, Kopernikusstr. 10, D-52074 Aachen, Germany*

[2] *Institute for Technical Physics and Materials Science, HUN-REN Centre for Energy Research, Konkoly-Thege Miklós út 29-33, H-1121 Budapest, Hungary*

[3] *Department of Physics and Astronomy, Uppsala University, Regementsvägen 10, S-75120 Uppsala, Sweden*

[4] *Plansee Composite Materials GmbH, Siebenbürgerstraße 23, D-86983 Lechbruck am See, Germany*

* Corresponding author; E-mail: kuemmerl@mch.rwth-aachen.de

**Abstract**

The composition-dependent thermal stability of $(Cr_{0.47\pm0.03}Al_{0.53\pm0.03})_z(O_yN_{1-y})_{1-z}$ thin films ($z = 0.475 \pm 0.015$) with O concentrations of $y = 0$, 0.15, and 0.40 is investigated up to 1200 °C and then compared to $(Ti_{0.56}Al_{0.44})_z(O_yN_{1-y})_{1-z}$. X-ray diffraction reveals a thermal stability limit of 1150 °C independent of the O concentration, as witnessed by the formation of decomposition products, namely h-$Cr_2N$ for $(Cr_{0.50}Al_{0.50})_{0.49}N_{0.51}$ and c-Cr for both $(Cr_{0.48}Al_{0.52})_{0.48}(O_{0.15}N_{0.85})_{0.52}$ and $(Cr_{0.44}Al_{0.56})_{0.46}(O_{0.40}N_{0.60})_{0.54}$. Based on transmission electron microscopy and elastic recoil detection analysis data, the thermal stability limit is extended to 1100 °C – 1150 °C for all films.

Chemical environment-dependent DFT calculations indicate that bond breaking limits the thermal stability. In (Cr,Al)N, N has the lowest activation energy for migration. Furthermore, the O vacancy formation energy is highest in (Cr,Al)(O,N). It has to be overcome to enable diffusion on the non-metal sublattice, which is necessary for forming decomposition products like w-AlN or c-Cr. However, once Cr-N bonds break, decomposition into h-$Cr_2N$ and subsequent c-Cr together with $N_2$ is triggered. This results in N evaporation, generating sufficient non-metal vacancies that greatly enhance diffusion and render the extensive vacancy formation energies for non-metals irrelevant. This reduction of the activation energy for mass transport on the non-metal sublattice to the migration barrier causes the similar thermal stability in $(Cr_{0.47\pm0.03}Al_{0.53\pm0.03})_z(O_yN_{1-y})_{1-z}$. In contrast, Al bonds break first without creating non-metal vacancies in (Ti,Al)(O,N). Thus, the high O vacancy formation energy in (Ti,Al)(O,N) significantly increases the thermal stability compared to (Ti,Al)N as well as the here investigated films.

# 1. Introduction

Metastable cubic transition metal aluminum nitride (c-(TM,Al)N) coatings are known for their excellent properties, including high wear resistance, hardness, and thermal stability [1–3]. Thermal stability is crucial for the performance of protective coatings in high-temperature environments, particularly during cutting processes where temperatures can surpass 1000 °C on the rake face at the cutting edge [4,5]. Compared to (Ti,Al)N, (Cr,Al)N shows a 10 at. % higher critical Al solubility before the wurtzite (w-)AlN phase forms in reactive magnetron sputtered films with an applied bias potential of -90 V [3]. The positive mixing enthalpy $\Delta H_m$ acts as the driving force for decomposition and is 200 % larger for $Ti_{0.5}Al_{0.5}N$ compared to $Cr_{0.5}Al_{0.5}N$, resulting in w-AlN formation between 800 °C to 900 °C, and 1100 °C to 1200 °C, respectively [3]. Thus, according to Rovere *et al.* [3], $Cr_{0.5}Al_{0.5}N$ exhibits higher thermal stability than (Ti,Al)N [3].

During annealing of c-(Cr,Al)N (space group $Fm\bar{3}m$, NaCl prototype), w-AlN (space group $P6_3mc$) is formed at grain boundaries and triple junctions via nucleation and growth [6]. Hence, the remaining c-(Cr,Al)N matrix becomes Cr-enriched, which destabilizes the Cr-N bonds [6]. Since the thermal stability of CrN is limited to 1050 °C at $10^5$ Pa [7], these Cr-enrichments lead to the decomposition into c-Cr and gaseous $N_2$ via intermediate hexagonal (h-)$Cr_2N$ [6]. This two-stage decomposition process is accompanied by a mass loss of up to 14.1 wt. % corresponding to 27 ± 2 at.% N for $(Cr_{0.5}Al_{0.5})N$ [3].

Mayrhofer *et al.* [8] studied the influence of Al on the thermal stability of $Cr_{1-x}Al_xN$ from c-CrN to w-AlN. The stability against N evaporation is enhanced as the Al content increases until the solubility limit is reached. At $x$ = 0.71, small amounts of w-AlN form below the detection threshold of X-ray diffraction (XRD), triggering decomposition and promoting N evaporation, which causes the lowest observed thermal stability [8]. At higher Al concentrations ($x \geq 0.83$), w-$Cr_{1-x}Al_xN$ forms, exhibiting greater thermal stability than c-$Cr_{1-x}Al_xN$, as w-AlN encapsulates c-CrN domains, slowing down Cr-N bond dissociation and enhancing overall thermal stability [8].

To further enhance properties such as thermal stability [9,10] or wear resistance [11,12], incorporating O into the non-metal sublattice to form (TM,Al)(O,N) is a promising approach. Therefore, the phase formation of (TM,Al)(O,N) (TM = Ti [13], Cr [12]) has been intensively studied through experiments and calculations [14].

When N atoms ($N^{3-}$) are substituted by O atoms ($O^{2-}$), the resulting charge imbalance is compensated by metal (e.g. $Cr^{3+}$) vacancy formation [13,15]. Furthermore, the incorporation of O enhances the ionic character of the bonding, resulting in longer and weaker TM-O and Al-O bonds in comparison to TM-N and Al-N [14]. Both the presence of metal vacancies and the weaker bonds contribute to a reduction in bulk modulus as the O content increases [13–15].

It has been shown that the thermal stability of ($Ti_{0.55}Al_{0.45}$)N can be significantly improved by adding O on the non-metal sublattice, forming (Ti,Al)(O,N) [9]. While w-AlN solid solution was formed in arc-deposited (Ti,Al)N thin films after vacuum annealing at 1000 °C, the w-AlN formation was shifted to 1300 °C in (Ti,Al)(O,N) with 15 at.% O. The decomposition of (Ti,Al)(O,N) into w-AlN and c-TiN solid solutions requires mobility not only on the metal sublattice, but also on the non-metal sublattice, while for (Ti,Al)N decomposition the activation of diffusion on the metal sublattice is sufficient [9]. Since the formation energy for non-metal vacancies is significantly larger in (Ti,Al)(O,N) than in (Ti,Al)N, higher temperatures are required to form w-AlN solid solution in (Ti,Al)(O,N), and thus the thermal stability is improved [9].

Delayed w-AlN formation from 900 °C to 1100 °C has also been reported for ($Ti_{0.42}Al_{0.58}$)N after adding O forming ($Ti_{0.41}Al_{0.59}$)$_{0.47}$($O_{0.18}N_{0.82}$)$_{0.53}$ [10]. However, the authors attributed the higher thermal stability to the increased diffusion activation energy of Al [10]. Additionally, c-Ti(O,N) was identified as a preferred decomposition product of (Ti,Al)(O,N) together with w-AlN. When the O content exceeded the solubility limit in TiN, nano-sized $Al_2O_3$ particles were found and suggested to cause a strengthening effect. It was argued that the hardness of $31.7 \pm 0.8$ GPa of as-deposited $(Ti_{0.41}Al_{0.59})_{0.47}(O_{0.18}N_{0.82})_{0.53}$ increases to $35.5 \pm 0.8$ GPa after annealing at 1200 °C, which overcompensates for the decrease in hardness caused by w-AlN formation [10].

Liu et al. [16] studied the influence of the O content on the thermal stability of $Cr_{1-x}Al_xN$ thin films. N-rich $(Cr_{0.77}Al_{0.23})_{0.50}N_{0.50}$, $(Cr_{0.76}Al_{0.24})_{0.49}(O_{0.08}N_{0.92})_{0.51}$, and $(Cr_{0.75}Al_{0.25})_{0.48}(O_{0.21}N_{0.79})_{0.52}$ films all showed comparable exothermic and endothermic features in the signal from differential scanning calorimetry (DSC). The two endothermic peaks at ~1100 °C and ~1420 °C were interpreted as the two-step decomposition of Cr-N bonds. First, from c-CrN to $h-Cr_2N$ and then to c-Cr, accompanied by $N_2$ release. These three (Cr,Al)(O,N) compositions exhibited very similar mass change curves, where the final mass reduction after annealing up to 1450 °C was ~18 % due to $N_2$ release upon decomposition. XRD data revealed that $(Cr_{0.77}Al_{0.23})_{0.50}N_{0.50}$ begins to decompose at 1000 °C into w-AlN and $h-Cr_2N$, with the nitride phase being completely consumed after annealing at 1200 °C, leading to the formation of c-Cr. O addition slightly increased the temperature for w-AlN formation by a maximum of ~100 °C. However, their decomposition into $h-Cr_2N$ and c-Cr appeared at identical temperatures of 1000 °C and 1200 °C, respectively, independent of the O content. For $(Cr_{0.75}Al_{0.25})_{0.48}(O_{0.21}N_{0.79})_{0.52}$, the c-(Cr,Al)(O,N) solid solution was already consumed at 1100 °C, which is 100 °C lower than for the films with lower O content. The hardness of these three films started to decrease from ~25 GPa to ~20 GPa after annealing at 1000 °C due to the formation of w-AlN. DSC, XRD, and hardness data all indicate that the thermal stability of $(Cr_{0.77}Al_{0.23})_{0.50}N_{0.50}$, $(Cr_{0.76}Al_{0.24})_{0.49}(O_{0.08}N_{0.92})_{0.51}$,

and $(Cr_{0.75}Al_{0.25})_{0.48}(O_{0.21}N_{0.79})_{0.52}$ is comparable and only minute differences can be observed. However, the authors [16] do not provide a detailed explanation of the mechanism limiting thermal stability.

Raab *et al.* [17] studied the thermal stability of arc-deposited (Cr,Al)(O,N) coatings from an $Al_{0.7}Cr_{0.3}$ target at different $N_2/O_2$ gas flow ratios. A marginal increase of thermal stability in He atmosphere has been reported with increasing O concentration, where the decomposition of CrN into h-$Cr_2N$ and subsequently c-Cr via $N_2$ release can be delayed from ~1090 °C to ~1150 °C. At N-contents below 10 at. %, CrN dissociates in a one-step process to Cr and $N_2$ without the intermediate formation of $Cr_2N$. $(Cr_{0.41}Al_{0.59})_{0.47}(O_{0.30}N_{0.70})_{0.53}$ showed peak temperatures in DSC at ~1150 °C and ~1180 °C, which correspond to the dissociation of CrN to h-$Cr_2N$ and c-Cr via $N_2$ release. The w-AlN phase first appeared at ~1000 °C in XRD. At 1290 °C, $Al_2O_3$ starts to appear. Interestingly, films with 16.0, 25.5, and 35.0 at. % O all show the same mass loss after annealing up to 1500 °C of around 8 %. However, the w-AlN phase fraction increased from 1130 °C to 1190 °C and 1500 °C, respectively.

Here, we investigate the influence of O addition on the thermal stability of $(Cr_{0.47\pm0.03}Al_{0.53\pm0.03})_z(O_yN_{1-y})_{1-z}$ thin films by experiments and quantum mechanical calculations. Thereby we explain the TM valence electron concentration-dependent differences between (Ti,Al)N and (Cr,Al)N as well as (Ti,Al)(O,N) and (Cr,Al)(O,N) on the thermal stability. Atom probe tomography (APT) and transmission electron microscopy (TEM) data reveal the changes in the local chemical composition at the nanometer scale during vacuum annealing up to 1200 °C. Combined with density functional theory (DFT) statistical envelope calculations of the activation energies for migration, the mechanisms limiting the thermal stability of $(Cr_{0.47\pm0.03}Al_{0.53\pm0.03})_z(O_yN_{1-y})_{1-z}$ are identified and compared to $(Ti_{0.44}Al_{0.56})_{0.45}(O_{0.27}N_{0.73})_{0.55}$ [9].

## 2. Experimental details

The $(Cr_{0.47\pm0.03}Al_{0.53\pm0.03})_z(O_yN_{1-y})_{1-z}$ thin films were deposited with reactive high power pulsed magnetron sputtering (HPPMS) in a CemeCon CC800/9 system using a $Cr_{0.4}Al_{0.6}$ target from Plansee Composite Materials GmbH, which has a surface area of 8.8 x 50 cm$^2$. Sapphire (0001) substrates were positioned at a distance of 4.5 cm from the target and heated to 420 °C. A reactive mixture of Ar/N$_2$/O$_2$ was utilized with flow rates for Ar and N$_2$ of 140 sccm, while the O$_2$ flow was adjusted between 0 to 5 and 7.5 sccm to achieve different O concentrations, leading to a total deposition pressure of 0.53 Pa. A Melec SIPP2000USB-10-500-S power supply, operated with a time-average power output of 2 kW as well as pulse on- and off-times of 50 and 1950 µs in the HPPMS mode, respectively. This resulted in a duty cycle of 2.5 % and a frequency of 500 Hz, along with a peak power density of 0.4 kW cm$^{-2}$. The obtained film thicknesses ranged from 1.9 to 2.1 µm after depositing for 40 min.

Subsequently, isothermal annealing was performed in a vacuum furnace with a base pressure of 1×10$^{-4}$ Pa. The annealing temperature was increased from 800 °C to 1200 °C with a holding time of 30 min. The heating and cooling rates were set to 10 K/min.

For structural analysis, a Bruker AXS D8 Discover General Area Detector Diffraction System was employed using Cu Kα (λ = 1.5406 Å) X-rays at a voltage of 40 kV and current of 40 mA in Bragg-Brentano geometry.

To determine the chemical composition, time-of-flight elastic recoil detection analysis (ToF-ERDA) was performed at the Tandem Accelerator Laboratory of Uppsala University [18]. Recoils were generated using a primary beam of 36 MeV $^{127}I^{8+}$ ions with time-energy coincidence spectra recorded in a detector telescope, equipped with thin carbon foils for ToF measurement and a gas detector for energy discrimination [19]. The measurement uncertainty of the O and N concentration was 3% relative of the deduced values with aliquot fractions for Cr and Al.

The mechanical properties of as-deposited and annealed films were assessed using quasistatic nanoindentation with a Hysitron TI-900 TriboIndenter. The indentation modulus and hardness were calculated following the Oliver and Pharr method [20]. A diamond indenter with Berkovich geometry, characterized by a Poisson's ratio of 0.07 and an elastic modulus of 1140 GPa was employed. Load-controlled measurements at 10 mN were conducted, causing maximum contact depths of 120 nm for the as-deposited samples, which is less than 7% of the film thickness. Each sample underwent 25 indentations, and the indenter area function was determined from measurements on fused silica. The measured indentation modulus values were converted to elastic modulus using a Poisson's ratio of 0.22, derived from linear interpolation between values of 0.154 for c-AlN [21] and 0.290 for c-CrN [22].

An FEI Helios NanoLab 600 dual-beam microscope with a Ga$^+$ focused ion beam (FIB) was employed for lamellae and APT specimen preparation according to standard lift-out techniques [23]. The same instrument was utilized for scanning transmission electron microscopy (STEM) imaging at an acceleration voltage of 30 kV and a current setting of 50 pA. The integrated EDAX Octane Elect energy dispersive X-ray spectroscopy (EDX) detector was used for line scans at settings of 12 kV and current levels reaching up to 1.6 nA.

The TEM investigations were performed in a C$_s$-corrected 200 kV Themis (Thermo Fisher) microscope with 0.8 Å resolution. Bright field (BF) and dark field (DF) images were obtained, and selected area electron diffraction (SAED) was performed. Additionally, high angle annular dark field (HAADF) images were captured in STEM mode, while elemental maps were generated from EDX spectrum images.

Local chemical composition measurements at the nanometer scale were conducted with laser-assisted APT using a CAMECA LEAP 4000X HR, equipped with an ultraviolet laser featuring a pulse width of 10 ps. The laser pulses were set to a frequency of 200 kHz and an energy of 10 pJ to enhance the measurement accuracy [24]. The base temperature was 60 K and the detection rate 0.5%.

## 3. Computational details

The vacancy formation energy ($E_f$) and diffusion migration energy barrier ($E_b$) calculations presented in this work are based on statistical envelope calculations [25–27]. To this end, the structure considered here for the calculations was the fcc B1 structure (space group $Fm\bar{3}m$, NaCl prototype), a 2 × 2 × 2 supercell with metal and non-metal sublattice, each comprising 32 atoms. To mimic the desired experimental composition of nitride and oxynitride, the metal sublattice was populated with Cr, Al, and vacancies. Additionally, the non-metal sublattice was populated with O and N for oxynitride. The random solid solution in the single-crystal models of (Cr,Al)N and (Cr,Al)(O,N) was obtained by distributing the metals and non-metals on the corresponding sublattice by the special quasi-random structure (SQS) method [28]. Accordingly, the composition $Cr_{0.24}Al_{0.24}N_{0.52}$ is represented by a metal sublattice containing 15 Al, 15 Cr atoms, and 2 vacancies, while the composition $Cr_{0.20}Al_{0.26}O_{0.22}N_{0.32}$ is represented by a metal sublattice with 15 Al, 12 Cr atoms, and 5 vacancies as well as a non-metal sublattice with 13 O and 19 N atoms.

To calculate $E_f$ and $E_b$, we used DFT as implemented in the Vienna Ab initio Simulation Package (VASP) [29,30]. Generalized gradient approximation (GGA) based on Perdew-Burke-Ernzerhof (PBE) [31] parametrization has been used to describe the electron-electron exchange and correlation interactions. The pseudo-potentials used for each of the elements in the calculations treat any semi-core states as valence, and the recommended potential is used as per the VASP website recommendation. The ion-electron interactions were described using the projector augmented wave method [32], with a plane-wave energy cut-off of 500 eV. The corresponding Brillouin zone was sampled with 7 × 7 × 7 Monkhorst–Pack $k$-point mesh [33]. The Methfessel–Paxton [34] smearing of 0.2 eV was applied. A convergence criterion of $10^{-6}$ eV (per supercell) was used for the total energy during the electronic self-consistency cycles and ionic relaxations during structural optimizations, and the total energy convergence of $10^{-4}$ eV (per supercell) was applied. Magnetism was considered only for Cr atoms. The comparison between starting magnetic moments of 1.0 $\mu_b$ and 3.0 $\mu_b$ led to the same resultant magnetic moment after structure relaxations. The ferromagnetic structure was considered with the lowest total energy after comparison among non-magnetic vs. antiferromagnetic and ferromagnetic, with the starting moment set to 3.0 $\mu_b$.

At first, a vacancy was created at each of the sites, resulting in 62 atoms for $Cr_{0.20}Al_{0.25}O_{0.22}N_{0.32}$ (30 metal atoms, 32 non-metal atoms) and 59 atoms for $Cr_{0.20}Al_{0.26}O_{0.22}N_{0.32}$ (27 metal atoms, 32 non-metal atoms), and $E_f$ was calculated separately as $E_f = E_i - E_0 + \mu_i$, where $E_i$ and $E_0$ are the total energy of the supercell with and without the vacancy, respectively, and $\mu_i$ is the chemical potential of the species $i$. $\mu_i$ was conventionally set equal to the energy-per-atom of the corresponding element in its stable solid structure (antiferromagnetic-bcc-Cr, fcc-Al, N$_2$ gas, O$_2$ gas). The envelope approach, in which the predicted variation in vacancy formation energies ($E_f$) and diffusion migration energy barriers ($E_b$) across atomic sites underscore their strong sensitivity to the local chemical environment, is explained elsewhere [25–27]. Subsequently, $E_b$ was calculated using the nudged elastic band (NEB) method [35] implemented in VASP.

The average migration activation energy ($\bar{E}_a$) is presented for each species and composition considered, where $\bar{E}_a = \bar{E}_f + \bar{E}_b$, with $\bar{E}_f$ denoting the average vacancy formation energy and $\bar{E}_b$ the corresponding average migration barrier [25–27].

## 4. Results and discussion

The chemical composition of the as-deposited $(Cr_{0.47\pm0.03}Al_{0.53\pm0.03})_z(O_yN_{1-y})_{1-z}$ films determined by ToF-ERDA is provided in Table 1. By increasing the $O_2$ flow during the deposition from 0 to 5 and 7.5 sccm, O concentrations of < 1 ($y = 0$), 8 ($y = 0.15$), and 21 at.% ($y = 0.40$) were obtained. Simultaneously, the non-metal to metal ratio is increased from 1.04 to 1.09 and 1.18 for $(Cr_{0.50}Al_{0.50})_{0.49}N_{0.51}$, $(Cr_{0.48}Al_{0.52})_{0.48}(O_{0.15}N_{0.85})_{0.52}$, and $(Cr_{0.44}Al_{0.56})_{0.46}(O_{0.40}N_{0.60})_{0.54}$ due to the formation of metal vacancies for charge balancing [13,15], which is in good agreement with literature reports [12,16,17].

*Table 1: Chemical composition of as-deposited $(Cr_{0.47\pm0.03}Al_{0.53\pm0.03})_z(O_yN_{1-y})_{1-z}$ films determined by ToF-ERDA.*

| Cr [at.%] | Al [at.%] | O [at.%] | N [at.%] | Notation |
|---|---|---|---|---|
| 24.4 ± 0.8 | 24.5 ± 0.8 | 0.4 ± 0.1 | 50.7 ± 1.5 | $(Cr_{0.50}Al_{0.50})_{0.49}N_{0.51}$ |
| 22.8 ± 0.7 | 25.0 ± 0.8 | 8.0 ± 0.2 | 44.2 ± 1.3 | $(Cr_{0.48}Al_{0.52})_{0.48}(O_{0.15}N_{0.85})_{0.52}$ |
| 20.4 ± 0.7 | 25.5 ± 0.9 | 21.4 ± 0.6 | 32.7 ± 1.0 | $(Cr_{0.44}Al_{0.56})_{0.46}(O_{0.40}N_{0.60})_{0.54}$ |

Figure 1 depicts the XRD patterns of the as-deposited films and after vacuum annealing up to 1200 °C. The diffraction peak at approximately 37.1° corresponds to the c-(Cr,Al)(O,N) (111) lattice plane and shifts towards 37.3° and 37.7° with increasing O concentration. This evolution is consistent with a reduction of the equilibrium volume due to the above-mentioned formation of metal vacancies, induced by O incorporation [13,15]. Furthermore, an increase in the Al concentration with rising O concentration is observed, which also results in a peak shift towards the c-AlN peak position. However, the (111) peak in $(Cr_{0.50}Al_{0.50})_{0.49}N_{0.51}$ is not between the c-CrN and c-AlN peak positions, but shifted to lower angles, suggesting that residual stress is present in the film. Furthermore, this peak shift might be explained by interstitial N incorporation [36] and/or the formation of Frenkel pairs [37], which is conceivable for the energetic deposition conditions using HPPMS. Based on the diffraction data, it is evident that single-phase cubic metastable $(Cr_{0.47\pm0.03}Al_{0.53\pm0.03})_z(O_yN_{1-y})_{1-z}$ solid solutions were formed for all compositions in the as-deposited state with a strong (111) texture.

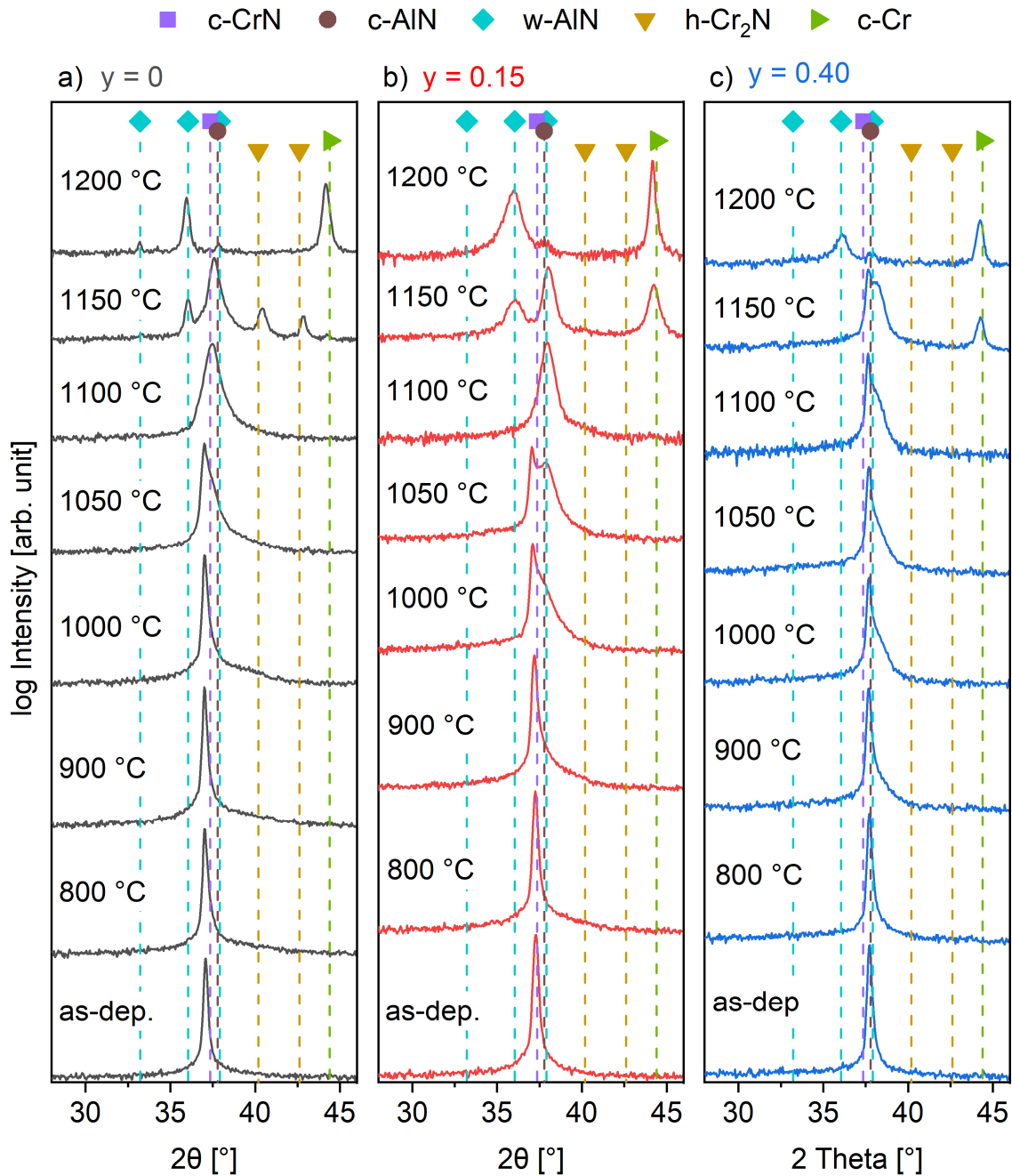

Figure 1: XRD patterns of (a) $(Cr_{0.50}Al_{0.50})_{0.49}N_{0.51}$, (b) $(Cr_{0.48}Al_{0.52})_{0.48}(O_{0.15}N_{0.85})_{0.52}$, and (c) $(Cr_{0.44}Al_{0.56})_{0.46}(O_{0.40}N_{0.60})_{0.54}$ thin films after vacuum annealing for 30 min at temperatures up to 1200 °C. Reference lines are according to the following PDF cards: c-CrN (01-083-5612, purple square), c-AlN (00-025-1495, brown pentagon), w-AlN (00-025-1133, blue rotated rectangle) h-$Cr_2N$ (00-035-0803, ochre triangle), and c-Cr (00-006-0694, green circle)

Up to 900 °C, the structure of the films seems unaffected, until peak broadening and shoulder formation start to appear at 1000 °C. Upon annealing, a peak shift to higher angles is observed until all films show the c-(Cr,Al)(O,N) peak at around 37.6° at 1100 °C, possibly caused by defect annihilation and stress relaxation [38]. Furthermore, spinodal decomposition also causes peak broadening and is observed by APT, especially in (Cr,Al)(O,N), and will be further analyzed after discussing the APT results in Figures 5-7. At 1150 °C, the films start to decompose, as indicated by the formation of decomposition products. While $(Cr_{0.50}Al_{0.50})_{0.49}N_{0.51}$ forms h-$Cr_2N$, the oxynitrides directly show the c-Cr peak, indicating that the N loss rate is significantly higher in oxynitride films compared to the nitride. Based on former reports, it is expected that CrN-rich regions decompose into h-$Cr_2N$ and further into c-Cr under $N_2$ release [16]. Furthermore, in $(Cr_{0.50}Al_{0.50})_{0.49}N_{0.51}$ and in $(Cr_{0.48}Al_{0.52})_{0.48}(O_{0.15}N_{0.85})_{0.52}$ w-AlN has formed additionally. At 1200 °C, all films show c-Cr together with w-AlN and a very small peak around 37.6° corresponding to the remaining matrix. The structural evolution suggests that the thermal stability of the $(Cr_{0.47\pm0.03}Al_{0.53\pm0.03})_z(O_yN_{1-y})_{1-z}$ films up to an O concentration of $y = 0.40$ is limited to 1150 °C, where the emergence of N-deficient phases is observed in XRD. The only difference is that the higher O concentration in $(Cr_{0.44}Al_{0.56})_{0.46}(O_{0.40}N_{0.60})_{0.54}$ delays the w-AlN formation by 50 °C to 1200 °C and thus leads to the formation of c-Cr before w-AlN. Liu *et al.* [16] reported that N-rich $(Cr_{0.76\pm0.01}Al_{0.24\pm0.01})_z(O_yN_{1-y})_{1-z}$ films annealed in Ar decompose into h-$Cr_2N$ at the same temperature as $(Cr_{0.77}Al_{0.23})N$, however, the w-AlN formation is delayed by ~100 °C for $(Cr_{0.76}Al_{0.24})_{0.49}(O_{0.08}N_{0.92})_{0.51}$ and $(Cr_{0.75}Al_{0.25})_{0.48}(O_{0.21}N_{0.79})_{0.52}$ compared to the nitride. This is in good agreement with our observations for $(Cr_{0.44}Al_{0.56})_{0.46}(O_{0.40}N_{0.60})_{0.54}$ but reported for films with higher Cr and lower O concentration [16].

While O addition leads to faster c-Cr formation without the intermediate h-$Cr_2N$ being detected in XRD, and the w-AlN formation is retarded by 50 °C for $y = 0.40$, the XRD results clearly exhibit that all $(Cr_{0.47\pm0.03}Al_{0.53\pm0.03})_z(O_yN_{1-y})_{1-z}$ films have the same thermal stability limit of 1150 °C when h-$Cr_2N$ or c-Cr is formed as a decomposition product.

Figure 2 shows the chemical composition measured by ToF-ERDA as a function of the annealing temperature. The chemical composition is constant within the measurement uncertainty until a significant drop in N concentration is first observed in $(Cr_{0.48}Al_{0.52})_{0.48}(O_{0.15}N_{0.85})_{0.52}$ ($y$ = 0.15) after annealing at 1100 °C, while the films with lower ($y$ = 0) and higher ($y$ = 0.40) O concentration reveal a significant reduction of the N concentration after annealing at 1150 °C. As observed in XRD, (Cr,Al)N decomposes into h-$Cr_2N$ and further into c-Cr under $N_2$ release due to breaking of Cr-N bonds. Therefore, the decrease in N concentration is an indicator for decomposition.

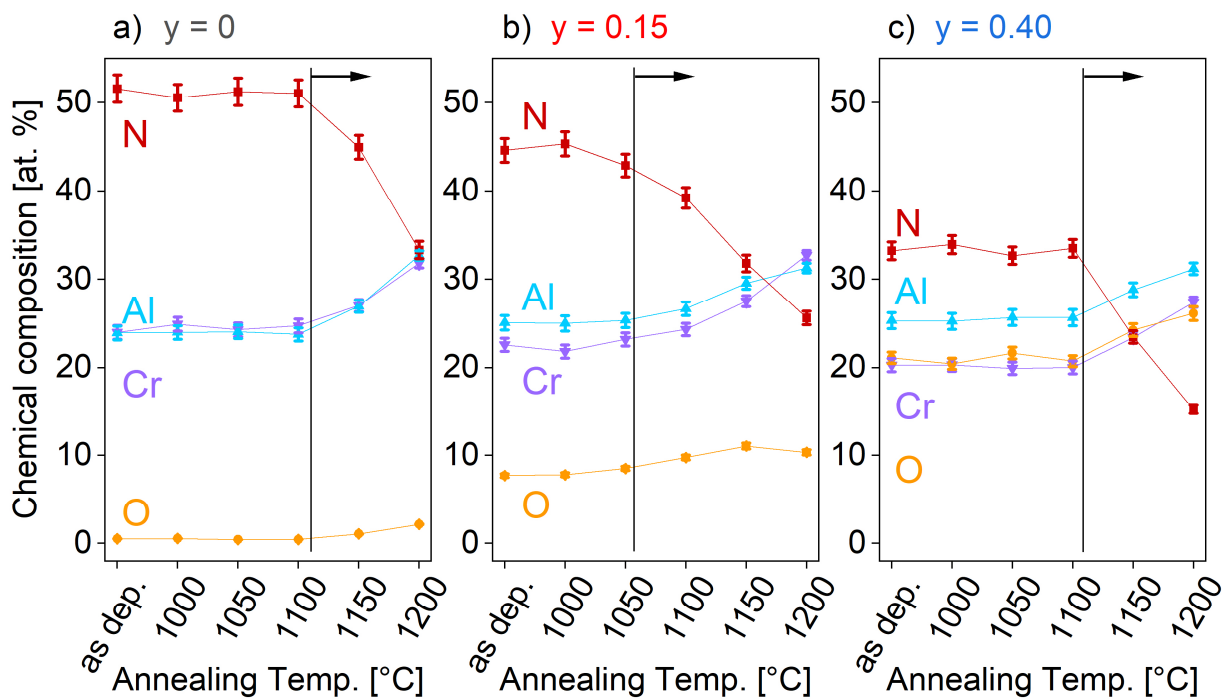

Figure 2: Chemical composition measured by ToF-ERDA as a function of the annealing temperature of (a) $(Cr_{0.50}Al_{0.50})_{0.49}N_{0.51}$, (b) $(Cr_{0.48}Al_{0.52})_{0.48}(O_{0.15}N_{0.85})_{0.52}$, and (c) $(Cr_{0.44}Al_{0.56})_{0.46}(O_{0.40}N_{0.60})_{0.54}$ thin films after vacuum annealing for 30 min at temperatures up to 1200 °C. The black lines and arrows indicate the temperature above which a significant reduction of the N concentration is observed.

$(Cr_{0.48}Al_{0.52})_{0.48}(O_{0.15}N_{0.85})_{0.52}$ is slightly less stable, inferred from the decrease in N concentration, and starts to decompose 50 °C earlier than $(Cr_{0.50}Al_{0.50})_{0.49}N_{0.51}$ and ($Cr_{0.44}Al_{0.56})_{0.46}(O_{0.40}N_{0.60})_{0.54}$. Therefore, consistent with XRD data, all films show a similar thermal stability limit of 1100 °C – 1150 °C, independent of the O concentration. Interestingly, the total mass loss after annealing at 1200 °C amounts to ~18 at.% N for all compositions. This comparable mass loss for N-rich (Cr,Al)(O,N) is in good agreement with literature reports [16,17].

Mechanical properties as a function of the annealing temperature are depicted in Figure 3. While the hardness of as-deposited films is similar with 33.3 ± 1.6 GPa, the elastic modulus is decreasing with increasing O concentration from 469 ± 16, to 456 ± 9, and 396 ± 6 GPa, respectively. Especially the higher O-containing $(Cr_{0.44}Al_{0.56})_{0.46}(O_{0.40}N_{0.60})_{0.54}$ film shows a 15 % lower modulus than $(Cr_{0.50}Al_{0.50})_{0.49}N_{0.51}$. This reduction can be explained by the enhanced ionic character of the bonding through O incorporation, resulting in longer and weaker Cr-O and Al-O bonds [14] as well as the presence of metal vacancies [13,15].

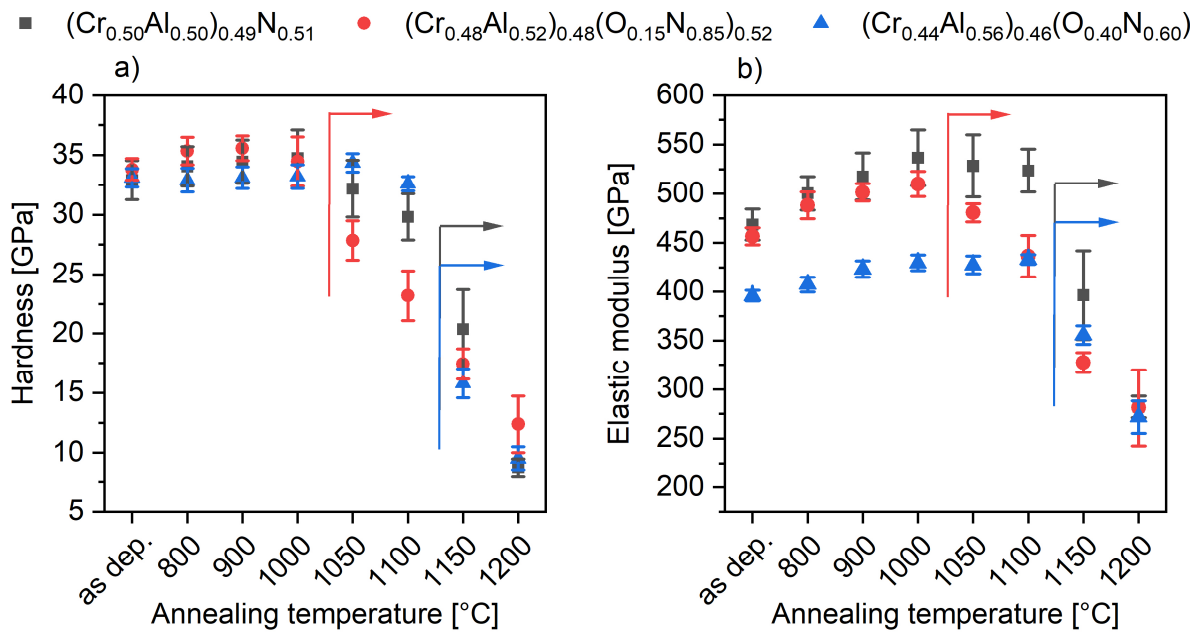

Figure 3: (a) Hardness and (b) elastic modulus as a function of the annealing temperature for $(Cr_{0.50}Al_{0.50})_{0.49}N_{0.51}$, $(Cr_{0.48}Al_{0.52})_{0.48}(O_{0.15}N_{0.85})_{0.52}$, and $(Cr_{0.44}Al_{0.56})_{0.46}(O_{0.40}N_{0.60})_{0.54}$ thin films after vacuum annealing for 30 min at temperatures up to 1200 °C. The arrows highlight the temperatures where the significant drop in mechanical properties appears first for the different compositions.

A notable reduction in both hardness and elastic modulus is first observed after annealing at 1050 °C for $(Cr_{0.48}Al_{0.52})_{0.48}(O_{0.15}N_{0.85})_{0.52}$. While there is a decrease in hardness for $(Cr_{0.50}Al_{0.50})_{0.49}N_{0.51}$ at the same temperature, this change is less pronounced and remains within the measurement uncertainty for hardness characterization, only becoming significant at 1150 °C. $(Cr_{0.44}Al_{0.56})_{0.46}(O_{0.40}N_{0.60})_{0.54}$ exhibits a substantial decline in mechanical properties after annealing at 1150 °C.

The XRD data, obtained in Bragg-Brentano geometry, in Figure 1 shows that the c-(Cr,Al)(O,N) solid solution peaks are shifted to lower angles than expected based on the c-CrN and c-AlN peak positions, indicating a compressive stress state of the as-deposited films. During annealing, a peak shift to higher angles is observed, possibly indicating defect annihilation and stress relaxation [38]. When compressive stress relaxation occurs, a decrease in elastic modulus is expected [39]. Accordingly, stress relaxation accounts for part of the observed modulus reduction. Furthermore, a decrease in hardness is expected when w-AlN is formed in (Cr,Al)N [40,41]. Even more severe might be the hardness decrease when c-Cr is formed, since the hardness of a pure Cr coating is only around ~2 GPa [42] to ~9 GPa [43]. For the formation of c-Cr, breaking of Cr-N bonds is a prerequisite, which results in $N_2$ release, which we capture during ToF-ERDA measurements. ToF-ERDA indicates a significant decrease in N concentration in $(Cr_{0.48}Al_{0.52})_{0.48}(O_{0.15}N_{0.85})_{0.52}$ for the sample annealed already at 1100 °C, while XRD reveals the formation of c-Cr and w-AlN at 1150 °C. Therefore, the observed reduction in hardness and elastic modulus in $(Cr_{0.48}Al_{0.52})_{0.48}(O_{0.15}N_{0.85})_{0.52}$ above 1050 °C is attributed to the formation of decomposition products and compressive stress relaxation. Whereby compressive stress relaxation might occur through defect annihilation as well as a reduction in molar volume upon decomposition under $N_2$ release. $(Cr_{0.50}Al_{0.50})_{0.49}N_{0.51}$ and $(Cr_{0.44}Al_{0.56})_{0.46}(O_{0.40}N_{0.60})_{0.54}$ films exhibit the reduction in mechanical properties at 1150 °C, concurrently with a decrease in N concentration and with the formation of $Cr_2N$ or c-Cr and w-AlN, respectively. Therefore, this decrease is attributed to the formation of soft phases like w-AlN and c-Cr and compressive stress relaxation upon decomposition and defect annihilation. The evolution of the microstructure upon vacuum annealing is depicted in Figure 4. All films show a dense, columnar microstructure in the as-deposited state. A decreasing film thickness with increasing O concentration is observed due to target poisoning of the reactively sputtered $O_2$ [44]. After annealing at 1000 °C, the films maintain their dense columnar microstructure, exhibiting neither signs of decomposition nor pore formation due to N evaporation. This observation aligns very well with the corresponding ERDA data presented in Figure 2, which show no evidence for N evaporation.

However, after annealing at 1050 °C, HAADF images of $(Cr_{0.50}Al_{0.50})_{0.49}N_{0.51}$ reveal the formation of dark regions along grain boundaries. In HAADF mode, dark contrast indicates areas with a lower atomic number, suggesting the presence of Cr-deficient and thus Al-rich regions that correspond to decomposed areas. However, only a slight decrease in hardness is observed, remaining within the measurement uncertainty. XRD analysis shows no signs of additional phases like w-AlN or h-$Cr_2N$, apart from a minor shift of peaks to higher angles and shoulder formation. In $(Cr_{0.48}Al_{0.52})_{0.48}(O_{0.15}N_{0.85})_{0.52}$, beginning from the surface, pores are formed at grain boundaries, indicating that decomposition and N evaporation have already started due to the bond breaking of Cr-N. These observations are in good agreement with the reduction of mechanical properties observed at 1050 °C for $(Cr_{0.48}Al_{0.52})_{0.48}(O_{0.15}N_{0.85})_{0.52}$. Also, ToF-ERDA data indicate a decrease in the average N-content. However, this decrease is still within the measurement uncertainty.

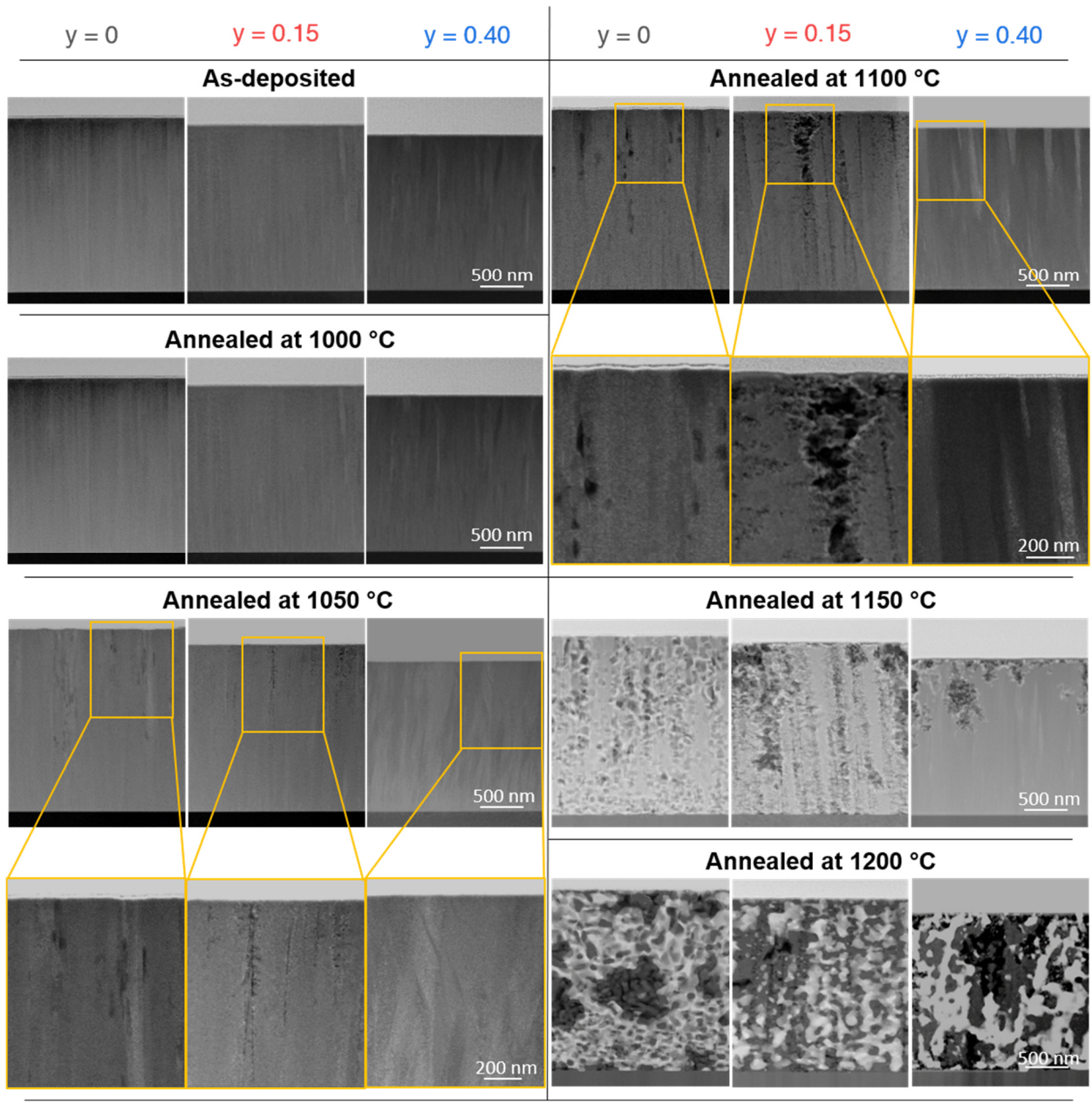

*Figure 4: Cross-sectional HAADF images in STEM mode showing the microstructure of $(Cr_{0.50}Al_{0.50})_{0.49}N_{0.51}$ – y = 0, $(Cr_{0.48}Al_{0.52})_{0.48}(O_{0.15}N_{0.85})_{0.52}$ – y = 0.15, and $(Cr_{0.44}Al_{0.56})_{0.46}(O_{0.40}N_{0.60})_{0.54}$ – y = 0.40 after vacuum annealing for 30 min at various temperatures up to 1200 °C.*

The $(Cr_{0.44}Al_{0.56})_{0.46}(O_{0.40}N_{0.60})_{0.54}$ film is still dense without any signs of decomposition at 1050 °C. When the temperature is raised to 1100 °C, the signs of decomposition and N-evaporation in the surface-near region along grain boundaries get more pronounced in $(Cr_{0.50}Al_{0.50})_{0.49}N_{0.51}$ and $(Cr_{0.48}Al_{0.52})_{0.48}(O_{0.15}N_{0.85})_{0.52}$, while $(Cr_{0.44}Al_{0.56})_{0.46}(O_{0.40}N_{0.60})_{0.54}$ still seems unaffected. As the temperature increases to 1150 °C, the decomposition process accelerates and propagates throughout the entire film thickness. Based on XRD in Figure 1, $(Cr_{0.50}Al_{0.50})_{0.49}N_{0.51}$ forms w-AlN as well as

h-$Cr_2N$, while w-AlN as well as c-Cr are formed in $(Cr_{0.48}Al_{0.52})_{0.48}(O_{0.15}N_{0.85})_{0.52}$. Also in $(Cr_{0.44}Al_{0.56})_{0.46}(O_{0.40}N_{0.60})_{0.54}$, significant decomposition and pore formation through $N_2$ release are observed for the first time at 1150 °C and are in good agreement with XRD data showing c-Cr formation. This decomposition clearly starts from the surface and evolves into the film, indicating that this process is driven by N release at the surface near region. At 1200 °C, all films are decomposed and only a small fraction of the (Cr,Al)(O,N) matrix remains.

To analyze the decomposition process further, near-atomic scale characterization was carried out by using APT and TEM, as presented in Figures 5-11. Cross-sectional three-dimensional APT characterization of $(Cr_{0.50}Al_{0.50})_{0.49}N_{0.51}$ is depicted in Figure 5. The local chemical composition analysis reveals the homogeneous distribution of Cr, Al, and N atoms at the nanometer scale in the as-deposited state (Figure 5a and b), with variations in composition within the count rate statistics. It has been demonstrated that laser-assisted APT is a precise tool to quantify the chemical composition of (Ti,Al)N and (Cr,Al)N thin films, but it lacks accuracy [24,45]. Thereby, the difference between the overall chemical composition of N measured by ERDA (51 at.%) and APT (45 at.%) can be rationalized.

After annealing at 1100 °C, compositional modulations of Cr and Al are mostly within the count rate statistics evident in Figure 5d). Locally larger variations in the Cr and Al contents are observed with maximum variations between 24 at.% and 32 at.% Cr as well as 21 at.% and 31 at.%. Al and an average modulation wavelength of 3 nm, but no significant spinodal decomposition or separate phase formation is observed.

When the temperature is further increased to 1150 °C, decomposition is observed in Figures 5e) and f). Alongside the remaining metastable cubic solid solution matrix, Cr-rich regions with up to 6 at.% Al and Al-rich regions with less than 10 at.% Cr develop, exhibiting a chemical composition similar to $Cr_2N$ and AlN, respectively. The measured decrease in N content within the Al-rich region compared to the solid solution may be affected by both the electric field strength [46] as well as by Cr-N bond breaking.

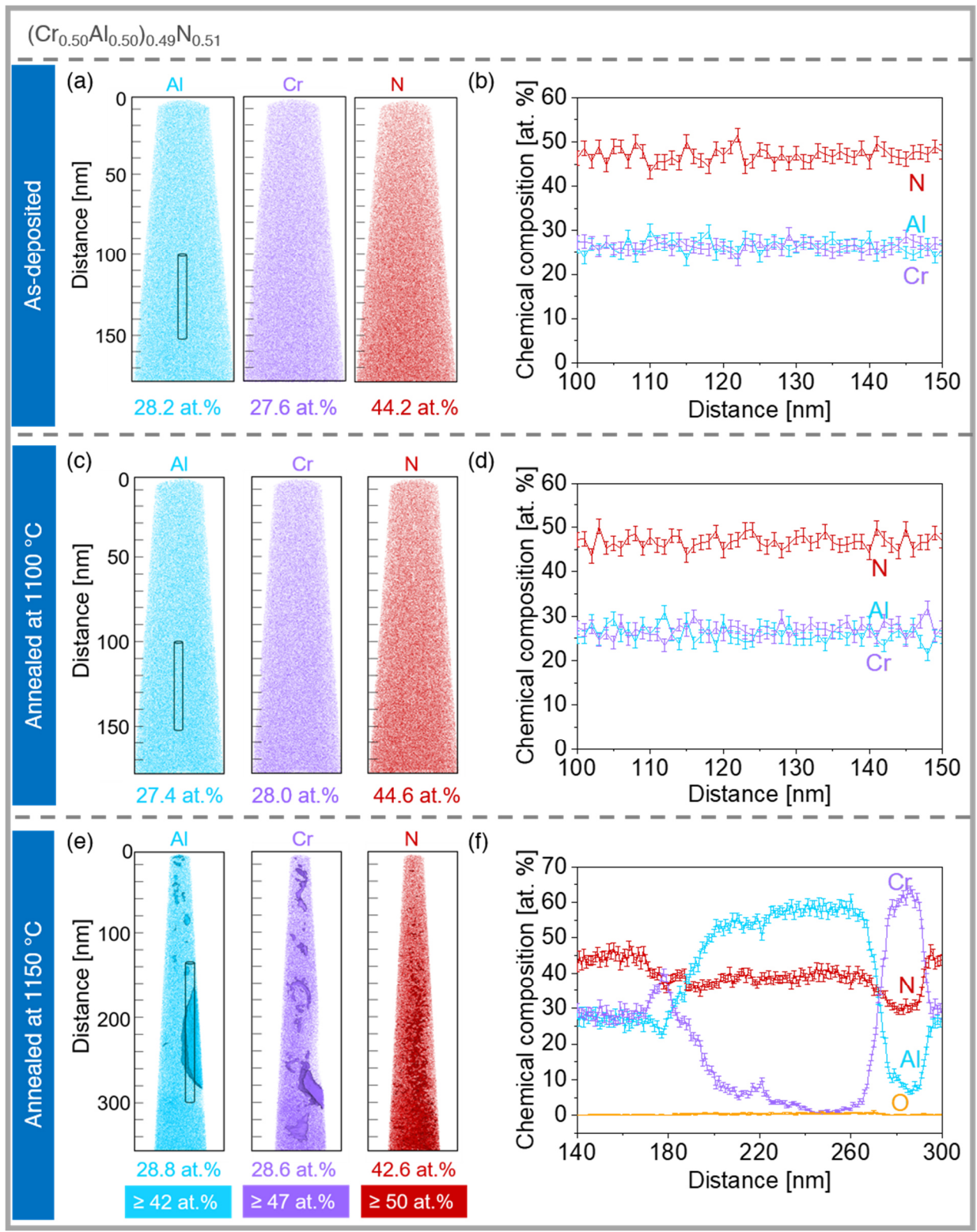

Figure 5: Cross-sectional APT characterization of (a) as-deposited $(Cr_{0.50}Al_{0.50})_{0.49}N_{0.51}$, as well as after vacuum annealing for 30 min at (c) 1100 °C and (e) 1150 °C with (b,d,f) respective composition profiles from the cylinders in (a,c,e). Isoconcentration surfaces of Cr ≥ 47 at.% and Al ≥ 42 at.% are highlighted for the atomic positions of Cr and Al in (e).

Evaporation of Al is characterized by a lower field, consequently, the local electric field is weaker, which reduces the ionization of neutral N-containing fragments and leads to underestimation [47]. However, Cr-N bond breaking and $N_2$ release may also contribute to the observed lower N concentration in the AlN solid solution, which based on the XRD data in Figure 1, corresponds to w-AlN.

When O is added forming $(Cr_{0.48}Al_{0.52})_{0.48}(O_{0.15}N_{0.85})_{0.52}$, a homogeneous distribution of Cr, Al, O, and N atoms is observed in the as-deposited state in Figure 6a) and b). After annealing at 1100 °C, periodic compositional modulations of Cr and Al are observed with maximum variations between 22 at.% and 29 at.% as well as 24 at.% and 33 at.% and a modulation wavelength of approximately 6 nm, respectively. These compositional modulations are considerably larger than the count rate statistics (approximately ±1 at.%) and consistent with spinodal decomposition together with the observed peak broadening in XRD data above 1000 °C in Figure 1.

At 1150 °C, decomposition has occurred, indicated by the Al- and Cr-rich regions depicted in Figure 6e) as isoconcentration surfaces in the reconstruction with more than 40 at.% of Al or 60 at.% of Cr, respectively. The composition profile in Figure 6f) indicates that, in addition to the presence of the solid solution, the formation of Al-rich regions with a similar composition to the Al-rich regions in $(Cr_{0.50}Al_{0.50})_{0.49}N_{0.51}$ (Figure 5f) have evolved and are identified as w-AlN by XRD as depicted in Figure 1. Furthermore, Cr-rich regions formed with up to 78 at.% Cr and 11 at.% of Al and N. XRD data in Figure 1 provide evidence that c-Cr is formed at 1150 °C without indications of h-$Cr_2$N formation. Therefore, this Cr-rich region might be a c-Cr solid solution. Interestingly, the O content in the Cr-rich regions is much smaller than the O concentration in the Al-rich regions and the metastable cubic solid solution, indicating a lower solubility of O in the Cr-rich phase compared to AlN.

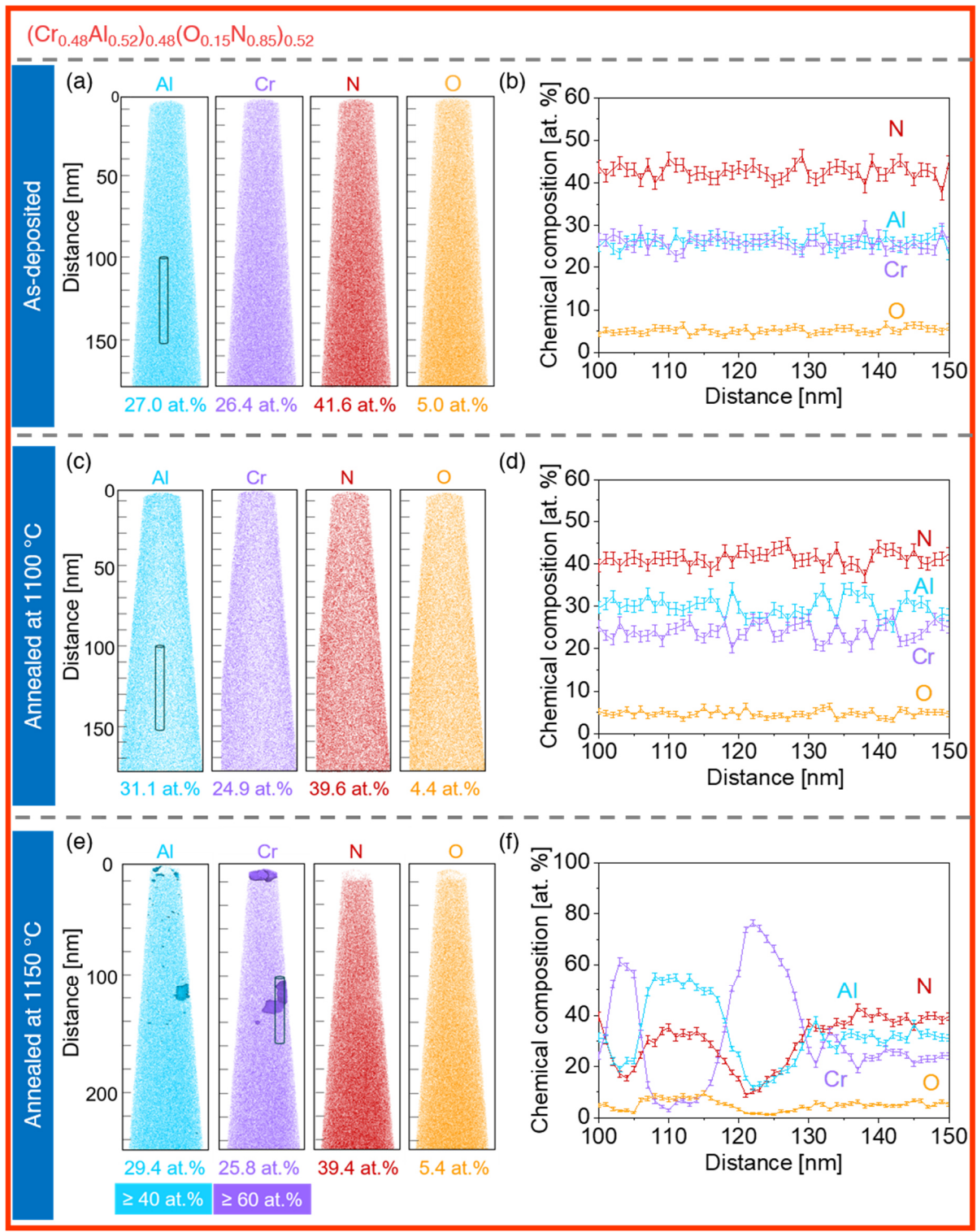

Figure 6: Cross-sectional APT characterization of (a) as-deposited $(Cr_{0.48}Al_{0.52})_{0.48}(O_{0.15}N_{0.85})_{0.52}$, as well as after vacuum annealing for 30 min at (c) 1100 °C and (e) 1150 °C with (b,d,f) respective composition profiles from the cylinders in (a,c,e). Isoconcentration surfaces of Cr ≥ 60 at.% and Al ≥ 40 at.% are highlighted for the atomic positions of Cr and Al in (e).

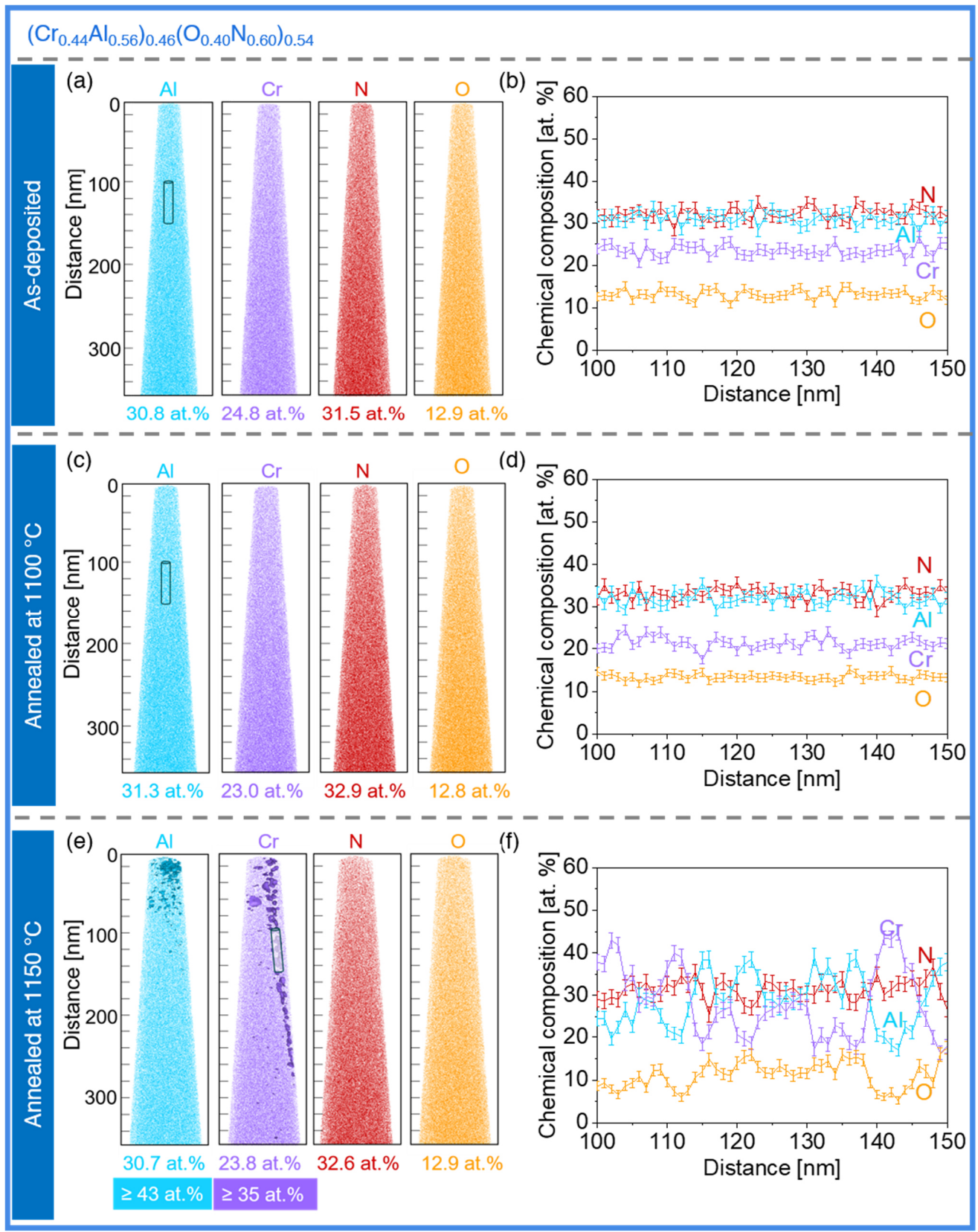

Figure 7: Cross-sectional APT characterization of (a) as-deposited $(Cr_{0.44}Al_{0.56})_{0.46}(O_{0.40}N_{0.60})_{0.54}$, as well as after vacuum annealing for 30 min at (c) 1100 °C and (e) 1150 °C with (b,d,f) respective concentration profiles from the cylinder in (a,c,e). Isoconcentration surfaces of Cr ≥ 35 at.% and Al ≥ 43 at.% are highlighted for the atomic positions of Cr and Al in (e).

For the as-deposited $(Cr_{0.44}Al_{0.56})_{0.46}(O_{0.40}N_{0.60})_{0.54}$ solid solution shown in Figure 7a) and b), variations in the Cr content are observed between 21 at.% and 27 at.% as well as 28 at.% and 34 at.% in Al and a modulation wavelength of approximately 5 nm. After annealing at 1100 °C (Figure 7c and d), the compositional modulations remain similar between 17 at.% and 24 at.% of Cr as well as 29 at.% and 36 at.% Al and modulation wavelength of approximately 5 nm. This indicates that even in the as-deposited state, the distribution of Cr and Al is not homogeneous, revealing surface diffusion initiated spinodal decomposition. After annealing at 1100 °C, bulk diffusion mediated spinodal decomposition is observed. As in the other films, Al- and Cr-rich regions have evolved after annealing at 1150 °C. However, based on the composition profile in Figure 7f), no phase separation, but enhanced spinodal decomposition is identified. APT data clearly support the notion that all three $(Cr_{0.47\pm0.03}Al_{0.53\pm0.03})_z(O_yN_{1-y})_{1-z}$ films exhibit a similar thermal stability, where no significant decomposition occurs below 1150 °C.

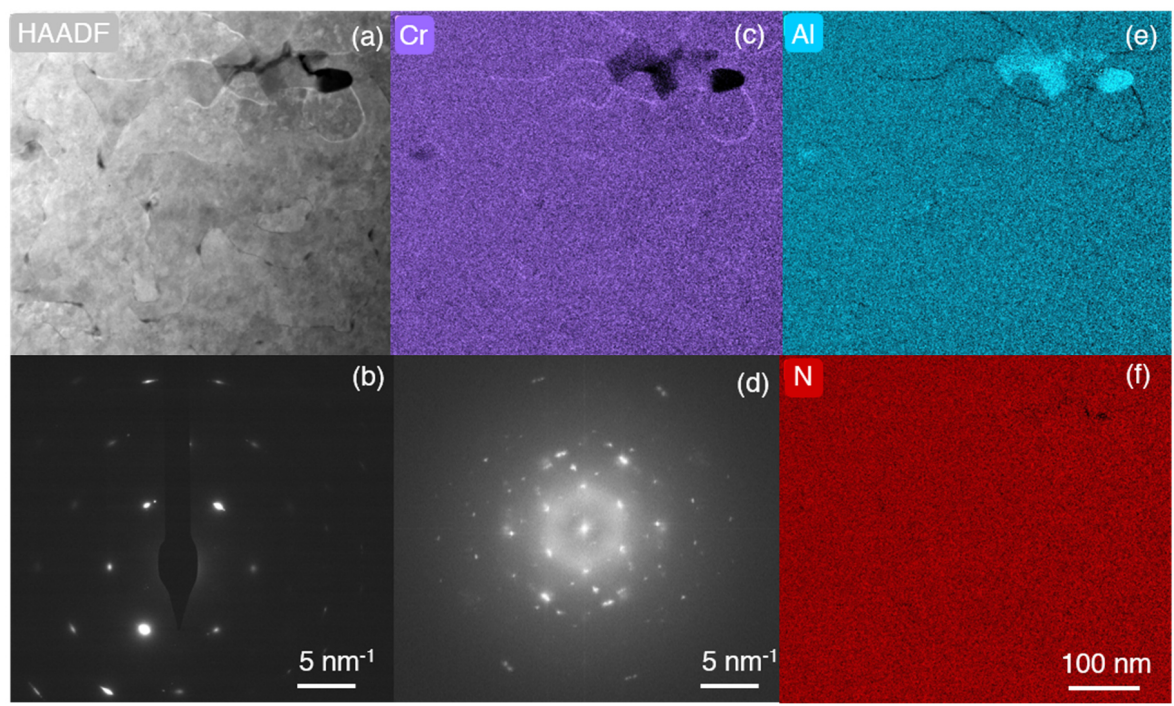

Figure 8: Plan-view TEM characterization of $(Cr_{0.50}Al_{0.50})_{0.49}N_{0.51}$ after vacuum annealing at 1100 °C for 30 min. (a) HAADF image. (b) SAED pattern. (d) FFT of a high resolution image from the Al-rich region (c,e,f) EDX elemental maps of Cr, Al, O, and N.

TEM was employed for spatially-resolved characterization of the nanostructure. A plan-view lamella of $(Cr_{0.50}Al_{0.50})_{0.49}N_{0.51}$ annealed at 1100 °C for 30 min was investigated by HAADF-STEM, and corresponding EDX maps are depicted in Figure 8. The EDX maps indicate that the majority of the $(Cr_{0.50}Al_{0.50})_{0.49}N_{0.51}$ matrix remains homogeneous. However, elemental separation occurs locally along grain boundaries and triple junctions, resulting in the formation of Al-rich and Cr-deficient grains. The SAED in Figure 8b) indicates a strongly (111) textured fcc pattern, and no evidence of w-AlN solid solution formation was observed in the TEM analysis, which aligns with the XRD data presented in Figure 1.

However, the fast Fourier transformation (FFT) of a high-resolution image from the Al-rich region in Figure 8d) is consistent with the formation of w-AlN, suggesting that the onset of decomposition occurs at 1100 °C at grain boundaries and triple junctions.

Figure 9 shows the TEM results for $(Cr_{0.48}Al_{0.52})_{0.48}(O_{0.15}N_{0.85})_{0.52}$ after vacuum annealing at 1100 °C for 30 min. In Figure 9a), a DF image is presented, and the orange circle marks a region where the SAED pattern (b) indicates only a strongly (111) textured fcc pattern. In the green area (c), in addition to the 1.44 Å (220) and 0.84 Å (422) diffractions of the cubic (CrAl)(ON) solid solution, additional weak diffractions can be observed at 2.67 Å and 1.55 Å, corresponding to w-AlN, and at 2.04 Å, corresponding to c-Cr, respectively. The weak diffraction at 1.18 Å is consistent with either w-AlN or c-Cr. The DF image (Figure 9a) taken with the 2.67 Å diffraction of w-AlN indicates that w-AlN has formed at grain boundaries and triple junctions in the bright areas of the dark field image. Figure 9 (d-i) shows the region, which is highlighted with a green circle in panel (a), at higher magnification and a c-Cr crystallite (marked by the yellow arrow) as well as w-AlN precipitates are present. The grain boundaries are O-rich, while Cr-enrichments surround the w-AlN precipitates. This indicates that precipitate formation originates at grain boundaries and triple junctions, while in the c-

(Cr,Al)(O,N) solid solution the original (111) texture is preserved. W-AlN forms at grain boundaries and triple junctions, which leads to the depletion of Al and thus the formation of Cr-rich regions. This Cr-enrichment destabilizes the bonds and leads to bond breaking, resulting in the decomposition into c-Cr under $N_2$ release.

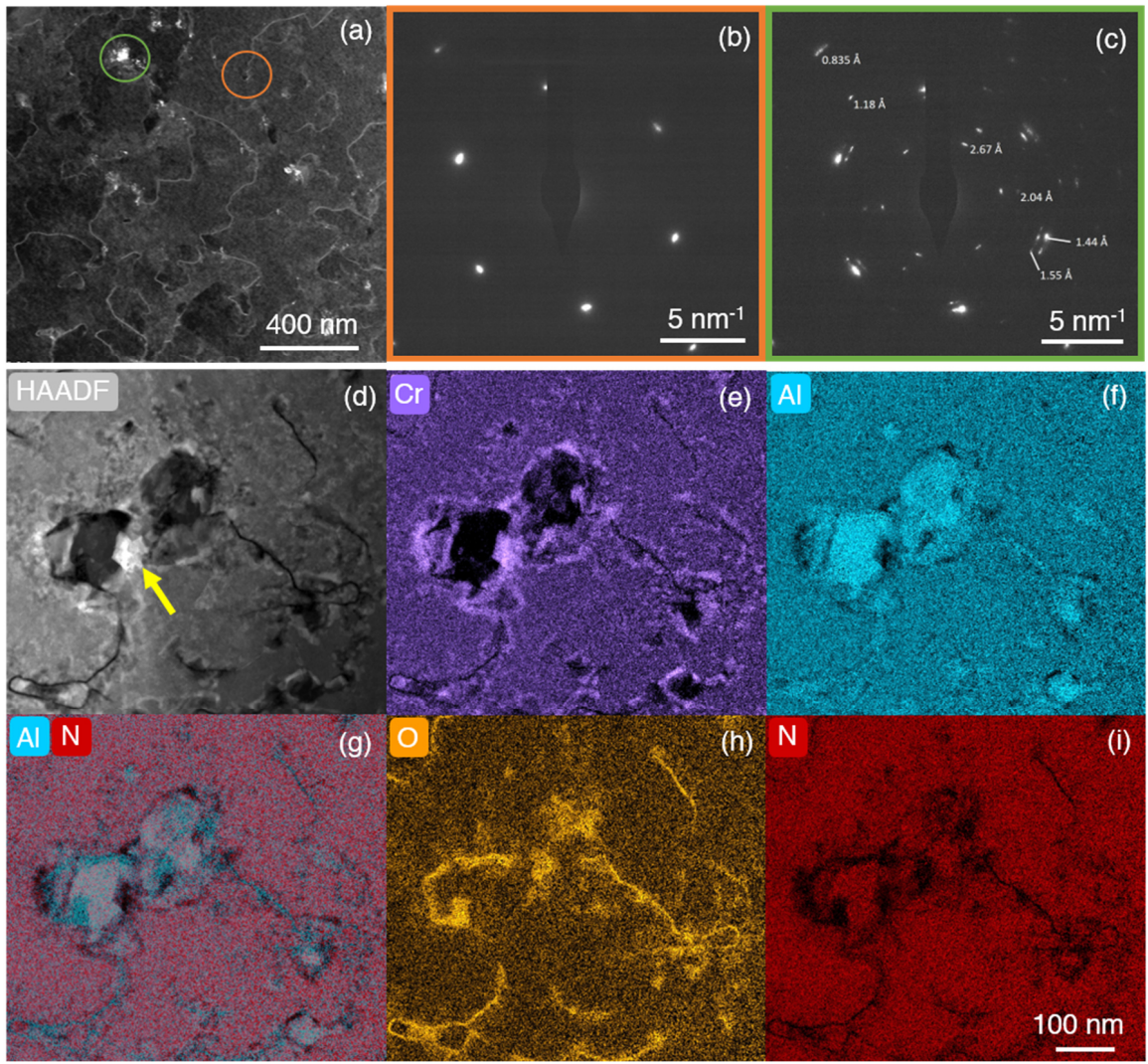

Figure 9: Plan-view TEM characterization of $(Cr_{0.48}Al_{0.52})_{0.48}(O_{0.15}N_{0.85})_{0.52}$ after vacuum annealing at 1100 °C for 30 min. (a) DF image taken with the diffraction of 2.67 Å corresponding to w-AlN. (b) SAED pattern from a non-transformed area as marked with an orange circle in (a). (c) SAED pattern from a transformed area as marked with a green circle in (a). (d) HAADF image (enlarged area of the green circle). (e)-(i) EDX elemental maps of Cr, Al, O, and N.

Figure 10 shows the TEM results of $(Cr_{0.44}Al_{0.56})_{0.46}(O_{0.40}N_{0.60})_{0.54}$. The solid solution exhibits uniform elemental distribution and no decomposition is observed. The SAED also indicates a strong (111) texture in agreement with XRD. In the high magnification, compositional modulations of Cr and Al are evident, consistent with spinodal decomposition as observed also in the APT data in Figure 7d).

The TEM investigations validate the findings from ERDA and nanoindentation, indicating that $(Cr_{0.48}Al_{0.52})_{0.48}(O_{0.15}N_{0.85})_{0.52}$ exhibits the lowest thermal stability, followed by $(Cr_{0.50}Al_{0.50})_{0.49}N_{0.51}$. In contrast, $(Cr_{0.44}Al_{0.56})_{0.46}(O_{0.40}N_{0.60})_{0.54}$ shows no signs of c-Cr or w-AlN formation at 1100 °C. However, after annealing at 1150 °C, all films form N-deficient phases like h-$Cr_2$N and c-Cr as discussed previously in Figure 1, which supports the observation of the thermal stability limit of 1100 °C – 1150 °C despite the significantly different O concentrations between the here investigated films.

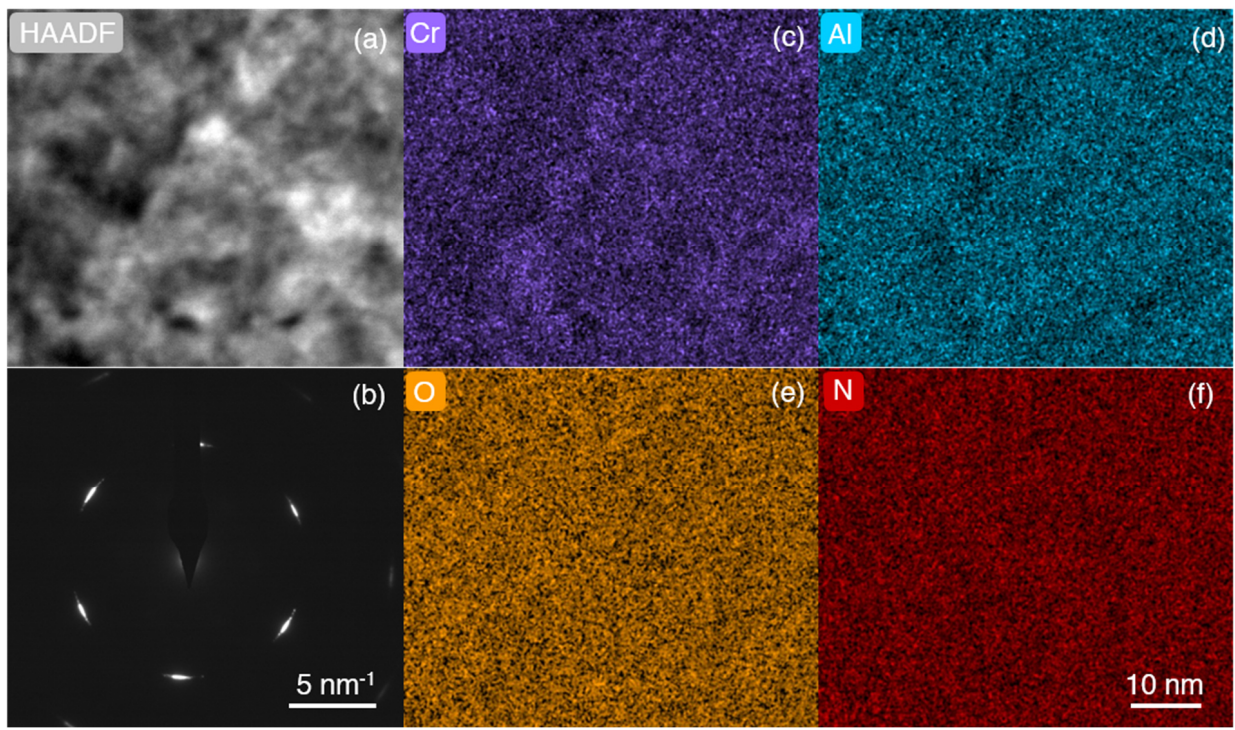

Figure 10: Plan-view TEM characterization of $(Cr_{0.44}Al_{0.56})_{0.46}(O_{0.40}N_{0.60})_{0.54}$ after vacuum annealing at 1100 °C for 30 min. (a) HAADF image. (b) SAED pattern. (c)-(f) EDX elemental maps of Cr, Al, O, and N.

The decomposition products are presented with an APT reconstruction of $(Cr_{0.44}Al_{0.56})_{0.46}(O_{0.40}N_{0.60})_{0.54}$ after annealing at 1200 °C in Figure 11. Based on the composition, three distinct phases can be identified. One phase consists primarily of c-Cr, with minimal presence of other elements. An Al-rich region has also developed, which can be recognized as w-AlN containing up to ~7 at.% O, based on the previous discussion and XRD analysis.

Additionally, the solid solution matrix remains locally, but is enriched in Al and O while being depleted in Cr and N compared to the as-deposited film illustrated in Figure 7a). This O-rich solid solution might transform to $Al_2O_3$ at higher temperatures and/or longer annealing times as described in [16] as a decomposition product of (Cr,Al)(O,N). The local compositional analysis at the nanometer scale also clearly demonstrates that the O content in the Cr phase with a maximum concentration of 98 at.% Cr and below 1 at.% O is significantly lower than in AlN with approximately 7 at.% O.

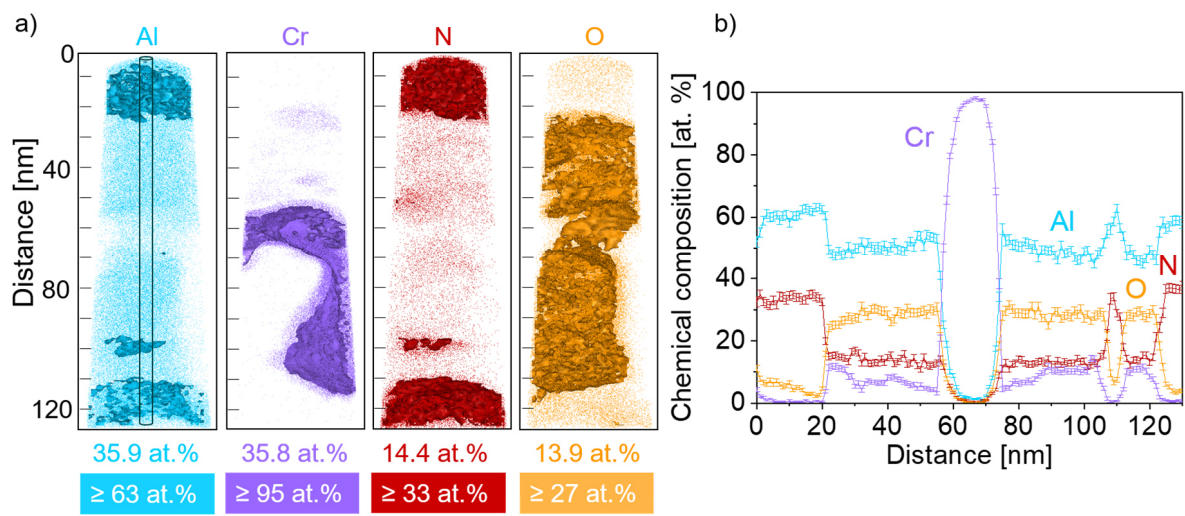

*Figure 11: APT characterization of $(Cr_{0.44}Al_{0.56})_{0.46}(O_{0.40}N_{0.60})_{0.54}$ after vacuum annealing for 30 min at 1150 °C. (a) Cr, Al, N, and O atomic positions with isoconcentration surfaces of Al ≥ 63 at.%, Cr ≥ 95 at.%, N ≥ 33 at.%, and, O ≥ 27 at.%. (b) Composition profile from the cylinder in (a).*

To investigate the mechanisms active during the thermal treatment, DFT envelope calculations were conducted and are presented in Figure 12. Ti- and Cr-based (TM,Al)N and (TM,Al)(O,N) are compared to explain the differences in thermal stability that depend on the TM valence electron concentration. Here, the envelopes capture the impact of different chemical environments encountered in the ternary nitride and quaternary oxynitride solid solutions during vacancy formation and diffusion migration. It was already shown for (Ti,Al)N that O addition significantly increases the thermal stability [9] and the chemical environment has an impact on the activation energy for migration in (Ti,Al)N$_x$ with different N stoichiometries[27]. Therefore, we analyzed activation energy ranges (envelopes) instead of the energy for only one selected migration pathway (single-point activation). The average activation energies consisting of the average vacancy formation energies plus average diffusion migration barriers with their standard deviations based on the different chemical environments of (Cr,Al)N and (Cr,Al)(O,N) are depicted in Figure 12 (b) and (d) in comparison to (Ti,Al)N (a) and (Ti,Al)(O,N) (c). The presented values are calculated as the sum of the average vacancy formation energy and the average diffusion migration energy.

The results demonstrate that in (Ti$_{0.50}$Al$_{0.50}$)$_{0.47}$N$_{0.53}$, Al exhibits the lowest average activation energy for migration of 6.6 ± 0.4 eV/atom. Consequently, the Al bond breaks first, making Al the fastest diffusing element and leading to the w-AlN formation observed experimentally [9]. In contrast, in (Cr$_{0.55}$Al$_{0.45}$)$_{0.46}$N$_{0.54}$, Al has the highest average activation energy for migration of 6.3 ± 0.6 eV/atom, while N has the lowest of 4.9 ± 0.8 eV/atom. Thus, bond breaking of N occurs first in this case. Once bond breaking of the Cr-N bond appears, decomposition into Cr$_2$N and subsequently into c-Cr under N$_2$ release is triggered. This results in the experimentally observed N evaporation.

Comparing the phase diagrams of TiN [48] and CrN [49] reveals that while c-TiN is stable up to ~42 at.% N before it starts to decompose into tetragonal (t-)Ti$_2$N, c-CrN exhibits a significantly smaller stability range and decomposes into h-Cr$_2$N when the N concentration is below approximately 49 at.%. Furthermore, c-TiN is with a formation energy of -1.9 eV/atom [50] significantly more stable than c-CrN (-0.7 eV/atom) [51], supporting the notion that the Cr-N bond is weaker and therefore breaks at lower temperatures compared to the Ti-N bond. Together with the fact that about 1 at.% of N understoichiometry in CrN readily causes decomposition, while TiN is able to accommodate about factor 8 larger N understoichiometry, the larger thermal stability of TiN compared to CrN can be understood.

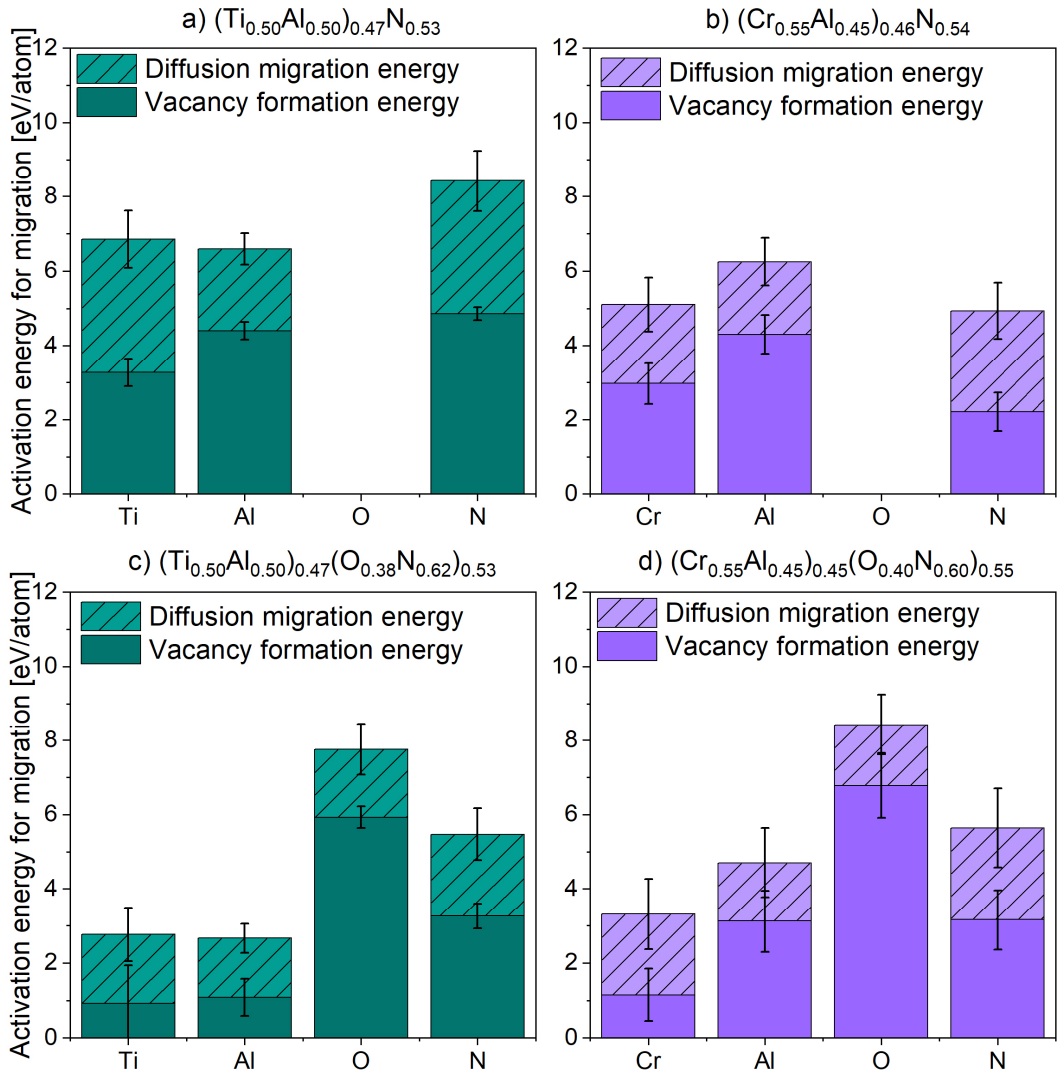

Figure 12: The average activation energy for migration plotted as the sum of the average vacancy formation energy and the average diffusion migration energy for (a) $(Ti_{0.50}Al_{0.50})_{0.47}N_{0.53}$ [9], (b) $(Cr_{0.55}Al_{0.45})_{0.46}N_{0.54}$, (c) $(Ti_{0.50}Al_{0.50})_{0.47}(O_{0.38}N_{0.62})_{0.53}$ [9], and (d) $(Cr_{0.55}Al_{0.45})_{0.45}(O_{0.40}N_{0.60})_{0.55}$. The standard deviations from the single-point calculations for the different chemical environments are plotted as well.

When O is added to the here studied nitrides, the activation energy for migration of the metals decreases significantly. This reduction likely arises first from the enhancement of the ionic bonding character upon O addition, causing longer and weaker bonds [14]. And second, in oxynitrides, a substantial amount of metal vacancies already exists due to maintaining electroneutrality, when N is replaced by O [13,15], leading to low metal vacancy formation energies. However, O diffusion requires the highest average activation energy for migration in $(Ti_{0.50}Al_{0.50})_{0.47}(O_{0.38}N_{0.62})_{0.53}$ as well as $(Cr_{0.55}Al_{0.45})_{0.45}(O_{0.40}N_{0.60})_{0.55}$ due to the high vacancy formation energies of 5.9 ± 0.3 and 6.8 ± 0.9 eV/atom, respectively. The vacancy formation energy for O is the largest and must be overcome to enable diffusion on the non-metal sublattice, which is the prerequisite for the formation of w-AlN or h-Cr$_2$N. This explains the significant improvement in thermal stability of $(Ti_{0.44}Al_{0.56})_{0.45}(O_{0.27}N_{0.73})_{0.55}$ [9].

However, in $(Cr_{0.55}Al_{0.45})_{0.45}(O_{0.40}N_{0.60})_{0.55}$, the Cr-N bond breaks first, as discussed previously. This bond breaking initiates the decomposition into Cr$_2$N and N$_2$ leading to N release, which creates a significant number of vacancies on the non-metal sublattice. Consequently, as N evaporates, numerous non-metal vacancies are formed, enabling mass transport on the corresponding sublattice. Thus, the O vacancy formation energy becomes irrelevant for vacancy formation on the non-metal sublattice since sufficient non-metal vacancies are generated during N release.

Therefore, the thermal stability in (Cr,Al)(O,N) is not limited by Al-bond breaking, leading to w-AlN solid solution formation as seen in (Ti,Al)(O,N) [9] but rather by the breaking of the Cr-N bond, resulting in N$_2$ release and creation of a sufficient amount of non-metal vacancies. These vacancies in turn enable diffusion on the non-metal sublattice, eliminating the need to overcome the high energy barrier associated with O vacancy formation energy as mobility enabler. Consequently, the experimentally observed O concentration independent thermal stability limit of 1100 °C – 1150 °C in (Cr,Al)(O,N) is rationalized based on chemical environment-dependent DFT calculations.

## 5. Conclusions

Single-phase cubic $(Cr_{0.47\pm0.03}Al_{0.53\pm0.03})_z(O_yN_{1-y})_{1-z}$ thin films with O concentrations of $y$ = 0, 0.15, and 0.40 and metal to non-metal ratios of $z$ = 0.49, 0.48, and 0.46 have been grown by reactive high-power pulsed magnetron sputtering, and the composition-dependence of the thermal stability has been investigated up to 1200 °C.

We report that O addition does not improve the thermal stability of (Cr,Al)N, contrary to (Ti,Al)N, where an increase of 300 °C in thermal stability upon O addition has been observed [9]. XRD data indicate identical thermal stability independent of the O content, as witnessed by the formation of N-deficient Cr-N phases at 1150 °C, namely h-$Cr_2N$ for $(Cr_{0.50}Al_{0.50})_{0.49}N_{0.51}$ and c-Cr for both $(Cr_{0.48}Al_{0.52})_{0.48}(O_{0.15}N_{0.85})_{0.52}$ and $(Cr_{0.44}Al_{0.56})_{0.46}(O_{0.40}N_{0.60})_{0.54}$. Higher O concentrations ($y$ = 0.40) delay w-AlN formation from 1150 °C to 1200 °C, whilealmost complete decomposition is observed at 1200 °C.

Complementary TEM and ToF-ERDA analyses reveal the onset of decomposition at 1100 °C in $(Cr_{0.48}Al_{0.52})_{0.48}(O_{0.15}N_{0.85})_{0.52}$. Hence, the oxygen concentration independent thermal stability limit lies within 1100 °C – 1150 °C. STEM images further reveal that the decomposition process is predominantly driven by N evaporation, initiated in surface-near regions.

The pronounced decrease in N concentration measured by ToF-ERDA can be rationalized by chemical environment-dependent DFT calculations, which show that N has the lowest activation energy for migration in (Cr,Al)N. The N bonds are the weakest and break first upon annealing. Once bond breaking of the Cr-N bond appears, decomposition into $Cr_2N$ and subsequently into c-Cr under $N_2$ release is triggered. This Cr-N bond breaking results in the experimentally observed formation of h-$Cr_2N$ and c-Cr accompanied by the reduction in N content, generating a large amount of non-metal vacancies. The CrN phase diagram further shows that c-CrN already becomes unstable and decomposes into h-$Cr_2N$ when the stoichiometry deviates by more than 1 at.% of N, while c-TiN accommodates the corresponding N understoichiometry without t-$Ti_2N$ formation.

When O is added to form (Cr,Al)(O,N), the formation of decomposition products such as w-AlN or c-Cr requires atomic mobility on both the metal and non-metal sublattices. DFT indicates that O exhibits the highest activation energy for migration due to the high O vacancy formation energy of 6.8 ± 0.9 eV/atom, whereas the vacancy formation energies of Cr, Al, and N are $\leq$ 4.3 ± 0.5 eV/atom. Thus, an increase in thermal stability would be expected. However, because the weak N bonds break first and $N_2$ evaporation creates sufficient non-metal vacancies, the large O vacancy formation energy becomes irrelevant to the effective thermal stability limit.

Hence, chemical environment-dependent DFT calculations explain the experimentally observed similarity in the thermal stability of (Cr,Al)N and (Cr,Al)(O,N). In contrast, for (Ti,Al)(O,N), Al–N bonds are the weakest and break first to form w-AlN without creating non-metal vacancies by $N_2$ release. In this case, the energy barrier for vacancy formation on the non-metal sublattice has to be overcome and the high energy required for O vacancy formation considerably increases the activation energy needed for mass transport on the non-metal sublattice. As a result, the thermal stability of (Ti,Al)(O,N) is significantly enhanced compared to that of (Ti,Al)N [9] as well as the here investigated Cr-based films. Finally, the observed divergence in thermal stability between (Ti,Al)(O,N) and (Cr,Al)(O,N) can be rationalized on the basis of chemical environment‑dependent DFT calculations.

**Acknowledgments**

This research was funded by the German Research Foundation (project 515702322). The TEM investigation was supported by the grant no. VEKOP-2.3.3-15-2016-00002 of the European Structural and Investment Funds. The Ministry for Culture and Science of the State of North Rhine Westphalia (MKW NRW) supported this work as part of the NHR funding. Computing time on the high-performance computer CLAIX at the NHR Center NHR4CES was provided with the project ID p0020883. Transnational ion beam analysis access has been supported by the RADIATE project under the Grant Agreement 824096 from the EU Research and Innovation program HORIZON 2020. Accelerator operation at Uppsala University has been supported by the Swedish research council VR-RFI (#2019-00191).

## References

[1] W. Münz, Titanium aluminum nitride films: A new alternative to TiN coatings, Journal of Vacuum Science & Technology A 4 (1986) 2717–2725. https://doi.org/10.1116/1.573713.

[2] H. Willmann, P.H. Mayrhofer, L. Hultman, C. Mitterer, Hardness evolution of Al–Cr–N coatings under thermal load, J. Mater. Res. 23 (2008) 2880–2885. https://doi.org/10.1557/JMR.2008.0366.

[3] F. Rovere, D. Music, S. Ershov, M.T. Baben, H.-G. Fuss, P.H. Mayrhofer, J.M. Schneider, Experimental and computational study on the phase stability of Al-containing cubic transition metal nitrides, J. Phys. D: Appl. Phys. 43 (2010) 035302. https://doi.org/10.1088/0022-3727/43/3/035302.

[4] T. Özel, T. Altan, Process simulation using finite element method — prediction of cutting forces, tool stresses and temperatures in high-speed flat end milling, International Journal of Machine Tools and Manufacture 40 (2000) 713–738. https://doi.org/10.1016/S0890-6955(99)00080-2.

[5] M.C. Shaw, ed., Metal cutting principles, 2nd ed, Oxford University Press, New York, 2005.

[6] H. Willmann, P.H. Mayrhofer, P.O.Å. Persson, A.E. Reiter, L. Hultman, C. Mitterer, Thermal stability of Al–Cr–N hard coatings, Scripta Materialia 54 (2006) 1847–1851. https://doi.org/10.1016/j.scriptamat.2006.02.023.

[7] T.B. Massalski, Binary Alloy Phase Diagrams, American Society for Metals, 1987.

[8] P.H. Mayrhofer, H. Willmann, A.E. Reiter, Structure and phase evolution of Cr–Al–N coatings during annealing, Surface and Coatings Technology 202 (2008) 4935–4938. https://doi.org/10.1016/j.surfcoat.2008.04.075.

[9] D.M. Holzapfel, D. Music, M. Hans, S. Wolff-Goodrich, D. Holec, D. Bogdanovski, M. Arndt, A.O. Eriksson, K. Yalamanchili, D. Primetzhofer, C.H. Liebscher, J.M.

Schneider, Enhanced thermal stability of (Ti,Al)N coatings by oxygen incorporation, Acta Materialia 218 (2021) 117204. https://doi.org/10.1016/j.actamat.2021.117204.

[10] J. Zhang, C. Hu, L. Chen, Y. Kong, Y. Du, P.H. Mayrhofer, Impact of oxygen content on the thermal stability of Ti-Al-O-N coatings based on computational and experimental studies, Acta Materialia 227 (2022) 117706. https://doi.org/10.1016/j.actamat.2022.117706.

[11] K. Tönshoff, B. Karpuschewski, A. Mohlfeld, T. Leyendecker, G. Erkens, H.G. Fuß, R. Wenke, Performance of oxygen-rich TiALON coatings in dry cutting applications, Surface and Coatings Technology 108–109 (1998) 535–542. https://doi.org/10.1016/S0257-8972(98)00637-9.

[12] H. Najafi, A. Karimi, P. Dessarzin, M. Morstein, Correlation between anionic substitution and structural properties in AlCr($O_xN_{1-x}$) coatings deposited by lateral rotating cathode arc PVD, Thin Solid Films 520 (2011) 1597–1602. https://doi.org/10.1016/j.tsf.2011.08.075.

[13] M. Hans, M. to Baben, D. Music, J. Ebenhöch, D. Primetzhofer, D. Kurapov, M. Arndt, H. Rudigier, J.M. Schneider, Effect of oxygen incorporation on the structure and elasticity of Ti-Al-O-N coatings synthesized by cathodic arc and high power pulsed magnetron sputtering, Journal of Applied Physics 116 (2014) 093515. https://doi.org/10.1063/1.4894776.

[14] S. Jin Rotert, D. Music, M. to Baben, J.M. Schneider, Theoretical study of elastic properties and phase stability of $M_{0.5}Al_{0.5}N_{1-x}O_x$ (M = Sc, Ti, V, Cr), Journal of Applied Physics 113 (2013) 083512. https://doi.org/10.1063/1.4793496.

[15] K.P. Shaha, H. Rueß, S. Rotert, M. to Baben, D. Music, J.M. Schneider, Nonmetal sublattice population induced defect structure in transition metal aluminum oxynitrides, Applied Physics Letters 103 (2013) 221905. https://doi.org/10.1063/1.4833835.

[16] Z.R. Liu, J.W. Du, L. Chen, Influence of oxygen content on structure, thermal stability, oxidation resistance, and corrosion resistance of arc evaporated (Cr, Al)N coatings, Surface and Coatings Technology 432 (2022) 128057. https://doi.org/10.1016/j.surfcoat.2021.128057.

[17] R. Raab, C.M. Koller, S. Kolozsvári, J. Ramm, P.H. Mayrhofer, Thermal stability of arc evaporated Al-Cr-O and Al-Cr-O/Al-Cr-N multilayer coatings, Surface and Coatings Technology 352 (2018) 213–221. https://doi.org/10.1016/j.surfcoat.2018.08.002.

[18] P. Ström, D. Primetzhofer, Ion beam tools for nondestructive in-situ and in-operando composition analysis and modification of materials at the Tandem Laboratory in Uppsala, J. Inst. 17 (2022) P04011. https://doi.org/10.1088/1748-0221/17/04/P04011.

[19] P. Ström, P. Petersson, M. Rubel, G. Possnert, A combined segmented anode gas ionization chamber and time-of-flight detector for heavy ion elastic recoil detection analysis, Review of Scientific Instruments 87 (2016) 103303. https://doi.org/10.1063/1.4963709.

[20] W.C. Oliver, G.M. Pharr, An improved technique for determining hardness and elastic modulus using load and displacement sensing indentation experiments, Journal of Materials Research 7 (1992) 1564–1583. https://doi.org/10.1557/JMR.1992.1564.

[21] B.D. Fulcher, X.Y. Cui, B. Delley, C. Stampfl, Hardness analysis of cubic metal mononitrides from first principles, Phys. Rev. B 85 (2012) 184106. https://doi.org/10.1103/PhysRevB.85.184106.

[22] A. Jain, S.P. Ong, G. Hautier, W. Chen, W.D. Richards, S. Dacek, S. Cholia, D. Gunter, D. Skinner, G. Ceder, K.A. Persson, Commentary: The Materials Project: A materials genome approach to accelerating materials innovation, APL Materials 1 (2013) 011002. https://doi.org/10.1063/1.4812323.

[23] K. Thompson, D. Lawrence, D.J. Larson, J.D. Olson, T.F. Kelly, B. Gorman, In situ site-specific specimen preparation for atom probe tomography, Ultramicroscopy 107 (2007) 131–139. https://doi.org/10.1016/j.ultramic.2006.06.008.

[24] H. Waldl, M. Hans, M. Schiester, D. Primetzhofer, M. Burtscher, N. Schalk, M. Tkadletz, Decomposition of CrN induced by laser-assisted atom probe tomography, Ultramicroscopy 246 (2023) 113673. https://doi.org/10.1016/j.ultramic.2022.113673.

[25] M. Hans, Z. Czigány, D. Neuß, J.A. Sälker, H. Rueß, J. Krause, G.K. Nayak, D. Holec, J.M. Schneider, Probing the onset of wurtzite phase formation in (V,Al)N thin films by transmission electron microscopy and atom probe tomography, Surface and Coatings Technology 442 (2022) 128235. https://doi.org/10.1016/j.surfcoat.2022.128235.

[26] G.K. Nayak, A. Kretschmer, P.H. Mayrhofer, D. Holec, On correlations between local chemistry, distortions and kinetics in high entropy nitrides: An ab initio study, Acta Materialia 255 (2023) 118951. https://doi.org/10.1016/j.actamat.2023.118951.

[27] G.K. Nayak, D. Holec, J.M. Schneider, Vacancy-concentration-dependent thermal stability of fcc-(Ti,Al)Nx predicted via chemical-environment-sensitive diffusion activation energies, (2025). https://doi.org/10.48550/arXiv.2510.16467.

[28] S.-H. Wei, L.G. Ferreira, J.E. Bernard, A. Zunger, Electronic properties of random alloys: Special quasirandom structures, Phys. Rev. B 42 (1990) 9622–9649. https://doi.org/10.1103/PhysRevB.42.9622.

[29] G. Kresse, J. Furthmüller, Efficient iterative schemes for *ab initio* total-energy calculations using a plane-wave basis set, Phys. Rev. B 54 (1996) 11169–11186. https://doi.org/10.1103/PhysRevB.54.11169.

[30] G. Kresse, J. Furthmüller, Efficiency of ab-initio total energy calculations for metals and semiconductors using a plane-wave basis set, Computational Materials Science 6 (1996) 15–50. https://doi.org/10.1016/0927-0256(96)00008-0.

[31] J.P. Perdew, A. Ruzsinszky, G.I. Csonka, O.A. Vydrov, G.E. Scuseria, L.A. Constantin, X. Zhou, K. Burke, Restoring the Density-Gradient Expansion for Exchange in Solids and Surfaces, Phys. Rev. Lett. 100 (2008) 136406. https://doi.org/10.1103/PhysRevLett.100.136406.

[32] P.E. Blöchl, Projector augmented-wave method, Phys. Rev. B 50 (1994) 17953–17979. https://doi.org/10.1103/PhysRevB.50.17953.

[33] H.J. Monkhorst, J.D. Pack, Special points for Brillouin-zone integrations, Phys. Rev. B 13 (1976) 5188–5192. https://doi.org/10.1103/PhysRevB.13.5188.

[34] M. Methfessel, A.T. Paxton, High-precision sampling for Brillouin-zone integration in metals, Phys. Rev. B 40 (1989) 3616–3621. https://doi.org/10.1103/PhysRevB.40.3616.

[35] D. Sheppard, R. Terrell, G. Henkelman, Optimization methods for finding minimum energy paths, J. Chem. Phys. 128 (2008) 134106. https://doi.org/10.1063/1.2841941.

[36] M. to Baben, L. Raumann, J.M. Schneider, Phase stability and elastic properties of titanium aluminum oxynitride studied by ab initio calculations, J. Phys. D: Appl. Phys. 46 (2013) 084002. https://doi.org/10.1088/0022-3727/46/8/084002.

[37] S. Karimi Aghda, D. Music, Y. Unutulmazsoy, H.H. Sua, S. Mráz, M. Hans, D. Primetzhofer, A. Anders, J.M. Schneider, Unravelling the ion-energy-dependent structure evolution and its implications for the elastic properties of (V,Al)N thin films, Acta Materialia 214 (2021) 117003. https://doi.org/10.1016/j.actamat.2021.117003.

[38] A. Hörling, L. Hultman, M. Odén, J. Sjölén, L. Karlsson, Thermal stability of arc evaporated high aluminum-content Ti1−xAlxN thin films, J. Vac. Sci. Technol. A 20 (2002) 1815–1823. https://doi.org/10.1116/1.1503784.

[39] D. Music, L. Banko, H. Ruess, M. Engels, A. Hecimovic, D. Grochla, D. Rogalla, T. Brögelmann, A. Ludwig, A. von Keudell, K. Bobzin, J.M. Schneider, Correlative plasma-surface model for metastable Cr-Al-N: Frenkel pair formation and influence of the stress state on the elastic properties, J. Appl. Phys. 121 (2017) 215108. https://doi.org/10.1063/1.4985172.

[40] M. Hirai, Y. Ueno, T. Suzuki, W. Jiang, C. Grigoriu, K. Yatsui, Characteristics of (Cr1-x, Alx)N Films Prepared by Pulsed Laser Deposition, Jpn. J. Appl. Phys. 40 (2001) 1056. https://doi.org/10.1143/JJAP.40.1056.

[41] M. Kawate, A. Kimura, T. Suzuki, Microhardness and lattice parameter of Cr1-xAlxN films, J. OF VACUUM SCIENCE & TECHNOLOGY A 20 (2002) 569–571. https://doi.org/10.1116/1.1448510.

[42] Y.-L. Su, K.-W. Cheng, J. Mayer, T.E. Weirich, A. Schwedt, Y.-F. Lin, Microstructure, phase transformation and hardness of nanometric Cr-Al multilayer coatings, Advances in Mechanical Engineering 7 (2015). https://doi.org/10.1177/1687814015589721.

[43] Z. Zeng, L. Wang, L. Chen, J. Zhang, The correlation between the hardness and tribological behaviour of electroplated chromium coatings sliding against ceramic and steel counterparts, Surface and Coatings Technology 201 (2006) 2282–2288. https://doi.org/10.1016/j.surfcoat.2006.03.038.

[44] M. Arif, C. Eisenmenger-Sittner, In situ assessment of target poisoning evolution in magnetron sputtering, Surface and Coatings Technology 324 (2017) 345–352. https://doi.org/10.1016/j.surfcoat.2017.05.047.

[45] M. Hans, J.M. Schneider, On the chemical composition of TiAlN thin films - Comparison of ion beam analysis and laser-assisted atom probe tomography with varying laser pulse energy, Thin Solid Films 688 (2019) 137251. https://doi.org/10.1016/j.tsf.2019.04.026.

[46] M. Hans, M. Tkadletz, D. Primetzhofer, H. Waldl, M. Schiester, M. Bartosik, C. Czettl, N. Schalk, C. Mitterer, J.M. Schneider, Is it meaningful to quantify vacancy concentrations of nanolamellar (Ti,Al)N thin films based on laser-assisted atom probe data?, Surface and Coatings Technology 473 (2023) 130020. https://doi.org/10.1016/j.surfcoat.2023.130020.

[47] M. Hans, J.M. Schneider, Electric field strength-dependent accuracy of TiAlN thin film composition measurements by laser-assisted atom probe tomography, New J. Phys. 22 (2020) 033036. https://doi.org/10.1088/1367-2630/ab7770.

[48] L. Wolff, G. Bastin, H. Heijligers, The titanium-nitrogen system, Solid State Ionics 16 (1985) 105–112. https://doi.org/10.1016/0167-2738(85)90031-1.

[49] S. Morita, H. Shimizu, Y. Sayama, Synthesis of chromium nitride powder by carbothermal nitriding, 2001.

[50] Materials Data on TiN by Materials Project, (2020). https://doi.org/10.17188/1208488.

[51] Materials Data on CrN by Materials Project, (2020). https://doi.org/10.17188/1196741.